\documentstyle[sprocl,epsf]{article}

\def\figlabel#1{\xdef#1{\thefigure}}
\def\fig#1{fig.~#1}
\def\figs#1{figs.~#1}
\def\figalign#1#2#3#4#5#6{
\begin{figure}
\centerline{
\hbox to 2truein{\vtop{\hsize=2truein\epsfxsize=5cm
\centerline{\epsfbox{#1} }
\caption[]{#3}
\figlabel{#2}
}}
\qquad\hbox to 2truein{\vtop{\hsize=2truein\epsfxsize=5cm
\centerline{\epsfbox{#4} }
\caption[]{#6}
\figlabel{#5}
}}
}
\end{figure}
}

\def\be{\begin{equation}}
\def\ee{\end{equation}}
\def\bea{\begin{eqnarray}}
\def\eea{\end{eqnarray}}

\begin{document}
\begin{flushright}
{ ~}\vskip -1in
CERN-TH/96-160\\
US-FT-32/96\\
hep-th/9606168
\end{flushright}

\title{SOFTLY BROKEN $N=2$ QCD \footnote{Based on a talk 
delivered by L. A.-G. at the Conference in honour of 
C. Itzykson ``The Mathematical Beauty of Physics".}}
\author{L. \'ALVAREZ-GAUM\'E}

\address{Theory Division, CERN, 1211 Geneva 23, Switzerland}

\author{M. MARI\~NO}

\address{Theory Division, CERN, 1211 Geneva 23, Switzerland\\ 
and \\Departamento de F\'\i sica de 
Part\'\i culas, Universidade de Santiago de Compostela,\\
E-15706 Santiago de Compostela, Spain}

\maketitle\abstracts{
We analyze the possible soft breaking of $N=2$ supersymmetric Yang-Mills 
theory with and without matter flavour preserving the analiticity 
properties of the Seiberg-Witten solution. 
We present the formalism for an arbitrary gauge group and obtain an 
exact expression for the effective potential. We describe in detail the 
onset of the confinement description and the vacuum structure for the 
pure $SU(2)$ Yang-Mills case and also some general features in the $SU(N)$ 
case. A general mass formula is obtained, as well as explicit results 
for the mass spectrum in the $SU(2)$ case.    
}

\section{Introduction and Conclusions.}
\setcounter{equation}{0}
In two remarkable papers \cite{swone,swtwo},
Seiberg and Witten obtained
exact information on the dynamics of $N=2$ supersymmetric gauge
theories in four dimensions with gauge group
$SU(2)$ and $N_{f} \le 4$
flavour multiplets. Their work was extended to
other groups in \cite{kl,klt,ds,groups}.
One of the crucial advantages of using
$N=2$ supersymmetry is that the low-energy
effective action in the Coulomb phase up to two derivatives
({\it i.e.} the K\"ahler potential,
the superpotential and the gauge kinetic function in
$N=1$ superspace language)
are determined in terms of a single holomorphic function called the
prepotential \cite{four}.
In references \cite{swone,swtwo},
the exact prepotential was determined using
some plausible assumptions and many consistency conditions.
For $SU(2)$ the solution is neatly presented by associating
to each case an elliptic curve together with
a meromorphic differential
of the second kind whose periods completely
determine the prepotential.
For other gauge groups \cite{groups} the solution
is again presented in terms
of the period integrals of a meromorphic differential
on a Riemann surface
whose genus is the rank of the group considered.
It was also shown in \cite{swone,swtwo} that by soft breaking $N=2$
down to $N=1$ (by adding a mass term for the adjoint $N=1$ chiral
multiplet in the $N=2$ vector multiplet) confinement follows
due to monopole condensation \cite{five}.

For $N=1$ theories exact results have also been obtained
\cite{none} using the holomorphy
properties
of the superpotential and the gauge kinetic function,
culminating in Seiberg's non-abelian
duality conjecture \cite{nadual}.

With all this new exact information it is also tempting to obtain
exact information about ordinary QCD. The obvious problem
encountered
is supersymmetry breaking. A useful avenue to explore is soft supersymmetry
breaking. The structure of soft supersymmetry breaking in $N=1$ theories has
been known for some time \cite{girar}. In \cite{softone,softwo}
soft breaking terms are used to explore $N=1$ supersymmetric QCD
(SQCD)
with gauge group $SU(N_c)$ and $N_f$ flavours of quarks,
and to extrapolate the exact results in \cite{none}
concerning the superpotential and the phase structure
of these theories in the absence of supersymmetry.
This leads to expected and unexpected predictions for
non-supersymmetric theories 
which may eventually be accessible to
lattice computations. In some cases however
for instance when $N_f \ge N_c$) it is known in the supersymmetric
case that the origin of moduli space is singular,
and therefore some of the assumptions made about the K\"ahler
potential
for meson and baryon operators are probably too strong.
Since the methods of \cite{swone,swtwo} provide us with
the effective action up to two derivatives,
the kinetic and potential term for all low-energy fields
are under control, and therefore in this paper we prefer
to explore in which way we can softly break
$N=2$ SQCD directly to $N=0$ while at the same time
preserving the analyticity properties of the Seiberg-Witten solution.
This is a very strong constraint and there is, essentially, only one
way to accomplish this task: we make the dynamical scale
$\Lambda$ of the $N=2$ theory a function of an
$N=2$ vector multiplet which is then frozen
to become a spurion whose $F$ and $D$-components break softly
$N=2$ down to $N=0$. If we want to interpret physically the
spurion, one can recall the string derivation of the
Seiberg-Witten solution in \cite{XI,XII} based on
type II-heterotic duality. In the field theory limit
in the heterotic side
(in order to decouple string and gravity loops)
the natural scaling is taken to be
$M {\rm e}^{iS} =\Lambda$, where $M$ is the Planck mass,
$S$ is the dilaton (in the low-energy theory $S=\theta/2\pi+ 4\pi
i/g^2$,
with $g$ the gauge coupling constant and
$\theta$ the CP-violating phase),
and $\Lambda$ the dynamical scale of the gauge theory
which is kept fixed while
$M \rightarrow \infty$ and $iS \rightarrow \infty$.
Since the dilaton sits in a vector multiplet of $N=2$
when the heterotic string is compactified on
$K3 \times T_{2}$, this is precisely the field
we want to make into a spurion, and 
procedure is compatible with the
Seiberg-Witten monodromies. In this way we obtain a
theory at $N=0$ with a more restricted structure
that those used in \cite{softone,softwo}.

As soon as the soft breaking terms
are turned on monopole condensation appears,
 and we get a unique ground state
(near the massless monopole point of
\cite{swone,swtwo}). Furthermore, in the Higgs
region we can compute
the effective potential, and we can verify that this potential
drives the theory towards the region where
condensation takes place. When the supersymmetry breaking parameter
is increased, the minimum displaces to the right along the
real $u$-axis. At the same time, the region in the $u$-plane
in which the monopole condensate is energetically-favoured expands.
Near the massless dyon point of \cite{swone,swtwo}, we find that dyon
condensation is energetically favourable but, unlike monopole condensation,
it is not sufficiently-strong an effect to lead to another minimum of the
effective potential. Eventually, when the soft supersymmetry breaking
parameter is made sufficiently large, the regions where monopole and dyon
condensation are favoured begin to overlap. At this point, it is clear that
our methods break down, and new physics is 
needed to describe the dynamics of
these mutually-nonlocal degrees of freedom.

  One advantage of this method
of using the dilaton spurion to softly break supersymmetry
from $N=2$ to $N=0$ is its universality.
It works for any gauge group and any number
of massive or massless quarks.
We work out the general structure of soft breaking by
the dilaton spurion in an arbitrary gauge group paying
special attention to the monodromies and the properties
of the spurion couplings, and we find the general features
of the vacuum structure for the case of $SU(N)$.
We also study the evolution of the mass eigenvalues 
in the case of the $SU(2)$ and also show 
in general that with this
soft breaking procedure there is a 
general sum rule satisfied by the masses of
all the multiplets.

The organization of this paper is as follows:
In section one we present the general formalism
for the breaking of supersymmetry due to a dilaton
spurion for a general gauge group, and we study  
the symplectic transformations of the various
quantities involved. The results agree with the general 
structure derived in \cite{XIII}
concerning the modification of the symplectic
transformations of special geometry in the
presence of background $N=2$ vector superfields. 
In section three we study the effective potential
and vacuum structure.  In section four we particularize
the formalism to the case of $SU(2)$ where the
analysis can be made more explicit.  In section
five we analyze in some detail the case for $SU(N)$
without hypermultiplets.  Finally in section six
we present a general mass sum rule for the general
case, and also obtain explicit results of the masses
in the $SU(2)$ case.  It is clear that for the moment
we cannot take the supersymmetry decoupling limit 
due to the fact that as the supersymmetry breaking
parameter increases we find that regions where mutually
non-local operators acquire vacuum expectation values
overlap.  This raises the fascinating issue that in order
to reach the real QCD limit we have to understand
the dynamics of the Argyres-Douglas phases \cite{ad}.

\section{Breaking $N=2$ with a dilaton spurion: general gauge group}
\setcounter{equation}{0}
In this section we present the generalization of 
the procedure introduced 
in \cite{soft} to $N=2$ Yang-Mills theories 
with a general gauge group $G$ of rank 
$r$ and massless matter hypermultiplets. 

The low energy theory description of the 
Coulomb phase \cite{swone} involves $r$ abelian 
$N=2$ vector superfields $A^i$, $i=1, \cdots, r$ 
corresponding to the unbroken 
gauge group ${U(1)}^r$. The holomorphic prepotential 
${\cal{F}} (A^i,\Lambda)$
 depends on 
the $r$ superfields $A^i$ and the dynamically 
generated scale of the theory, 
$\Lambda$. The low energy effective lagrangian 
takes the form (in $N=1$ notation)
\cite{swone}:
\be
{\cal L}={1 \over 4 \pi}{\rm Im}
\Bigl[ \int d^4 \theta {\partial {\cal F}\over
\partial A^i} {\overline A}^i + {1\over 2}
\int d^2 \theta {\partial ^2 {\cal F} \over
\partial A^i \partial A^j} W^i_{\alpha}W^{\alpha  j} \Bigr],
\label{lagr}
\ee
We define the dual variables, as in the $SU(2)$ case, by
\be
a_{D,i} \equiv {\partial {\cal F} \over \partial a^i}.
\label{two}
\ee
The K\"ahler potential and effective couplings 
associated to (\ref{lagr}) are:    
$$
K(a, {\bar a})={1 \over 4 \pi}{\rm Im} a_{D,i}{\bar a}^i,
$$
\be
\tau_{ij}= {\partial ^2 {\cal F} \over \partial a^i
\partial a^j}, 
\label{kahler}
\ee
and the metric of the moduli space is given accordingly by:
\be
(ds)^2={\rm Im}{\partial ^2 {\cal F} \over \partial a^i
\partial a^j}da^i d{\overline a}^j.
\label{metric}
\ee 
We introduce now a complex space ${\bf C}^{2r}$ with elements of the form
\be
v=\left(\begin{array}{c}a_{D,i} \\
                    a^i \end{array}\right).
\label{vector}
\ee
The metric (\ref{metric}) can then be written as
\bea
(ds)^2 &=& -{i \over 2}\sum_{i} (da_{D,i}d{\overline a}^i
-d {\overline a}_{D,i}da^i)
\nonumber\\
& \nonumber \\
&=& -{i \over 2} \left(\matrix{da_{D,i}& da^i}\right)
\left(\matrix{0& {\bf 1} \cr
               -{\bf 1}&0\cr}\right) 
\left(\begin{array}{c}d{\overline a}_{D,i} \\
                    d{\overline a}^i \end{array}\right),
\label{metmat}
\eea
which shows that the transformations of $v$ 
preserving the form of the metric 
are matrices $\Gamma \in Sp(2r, {\bf Z})$. 
They verify $\Gamma^{\rm T} \Omega 
\Gamma=\Omega$, 
where $\Omega$ is the $2r \times 2r$ matrix 
appearing in (\ref{metmat}), and
can be written as:
\be
\left(\matrix{A& B \cr
               C&D\cr}\right) 
\label{simp}
\ee
where the $r \times r$ matrices $A$, $B$, $C$, $D$ satisfy:
\be
A^{\rm T}D-C^{\rm T}B= {\bf 1}_r, \,\,\,\ 
A^{\rm T}C=C^{\rm T}A, \,\,\,\ 
B^{\rm T}D=D^{\rm T}B.
\label{simple}
\ee
The vector $v$ transforms then as:
\be
\left(\begin{array}{c}a_{D} \\
                    a \end{array}\right) \rightarrow \Gamma
\left(\begin{array}{c}a_D\\
               a \end{array}\right) =\left(\begin{array}{c}Aa_D+Ba\\
               Ca_D+Da \end{array}\right).
\label{trans}
\ee
From this we can obtain the modular transformation 
properties of the prepotential
 ${\cal F}(a^i)$ (see \cite{sonn}). Since
\bea
{\partial {\cal F}_{\Gamma} \over \partial a^k} &=&
{\partial a^i_{\Gamma} \over \partial a^k}
{\partial {\cal F}_{\Gamma}\over \partial a^i_{\Gamma}}=
\Bigl( C^{ip} \tau_{pk}+D^i_k \Bigr)
\Bigl(A_i^j a_{D,j} +B_{ij}a^j \Bigr)\nonumber \\
& \nonumber \\
&=& (D^{\rm T} B)_{kj}a^j+(D^{\rm T}A)_k^j{ \partial {\cal F} \over \partial 
a^j} + 
(C^{\rm T} B)^p_j {\partial a_{D,p} \over \partial a^k} a^j\nonumber\\
& \nonumber\\
&+& (C^{\rm T}A)^{pj} {\partial a_{D,p} \over \partial a^k}a_{D,j},
 \label{xxiii}
\eea
using the properties (\ref{simple}) of the 
symplectic matrices we can integrate 
(\ref{xxiii}) to obtain:
\bea
{\cal F}_{\Gamma}&=& {\cal F} +
{1 \over 2} a^k(D^{\rm T} B)_{kj}a^j+ {1 \over 2} a_{D,k} 
(C^{\rm T}A)^{pj}a_{D,j}\nonumber\\
& \nonumber\\
&+& a^k (B^{\rm T}C)_k^j a_{D,j}.
\label{premono}
\eea
Starting with (\ref{premono}) we can prove that the quantity 
${\cal F}-{1/2}\sum_{i}a^ia_{D,i}$ 
is a monodromy invariant, 
and evaluating it asymptotically, one obtains the relation 
\cite{matone,sonn,ey}:
\be
{\cal F}-{1 \over 2}\sum_{i}a^i a_{Di}=-4 \pi i b_1 u,
\label{mato}
\ee
where $b_1$ is the coefficient of the 
one-loop beta function (for $SU(N_c)$ with $N_f$ 
hypermultiplets in the fundamental representation, 
$b_1=(2N_c-N_f)/16 {\pi}^2$) and 
$u=\langle {\rm Tr} \phi^2 \rangle$. With the normalization 
for the electric charge used in \cite{swtwo} and 
\cite{ds}, the r.h.s. of 
(\ref{mato}) is $-2 \pi i b_1 u$. 

As in the $SU(2)$ case, presented in \cite{soft}, 
we break $N=2$ supersymmetry down to 
$N=0$ by making the dynamical scale $\Lambda$ a 
function of a background vector 
superfield $S$, $\Lambda = {\rm e}^{iS}$. 
This must be done in such a way 
that $s$, $s_D= \partial {\cal F} /\partial s$ 
be monodromy invariant. To see this, 
we will derive a series of relations analogous 
to the ones in the $SU(2)$ case \cite{soft}, 
starting with the following expression for the 
prepotential in terms of local 
coordinates:
\be
{\cal F}=\sum_{ij}a^i a^j f_{ij}(a^l/\Lambda),
\label{expansion}
\ee
where we take $f_{ij}=f_{ji}$. We define 
now a $(r+1)\times (r+1)$ matrix of couplings 
including the dilaton spurion $a^0=s$:
\be
\tau_{\alpha \beta}={{\partial}^2 
{\cal F}\over \partial a^{\alpha} a^{\beta}}.
\label{dilcouplings}
\ee
Greek indices $\alpha$, $\beta$ go from $0$ to $r$, 
and latin indices $i$, $j$ from
 $1$ to $r$. We obtain:
$$
a_{D,k}=2\sum_i a^if_{ik} + {1 \over \Lambda} \sum_{ij}a^i a^j f_{ij,k},
\nonumber
$$
$$
\tau_{ij}=2f_{ij}+{2 \over \Lambda}\sum_ka^k(f_{ik,j}+f_{jk,i})+
{1 \over \Lambda^2} a^ka^lf_{kl,ij},
\nonumber
$$
$$
\tau_{0i}=-{i\over \Lambda}\sum_{jk}a^ja^k(2f_{ij,k}+
f_{jk,i})-{i \over {\Lambda}^2}
\sum_{jkl}a^ja^ka^lf_{jk,li},
\nonumber
$$
\be
\tau_{00}=-{1 \over \Lambda}\sum_{ijk}a^ia^ja^kf_{ij,k}-
{1\over \Lambda^2}\sum_{ijkl}a^ia^ja^ka^lf_{ij,kl},
\label{sumas}
\ee
and the dual spurion field is given by:
\be
s_{D}={\partial {\cal F} \over \partial s}=
-{i \over \Lambda} \sum_{ijk}a^ia^jf_{ij,k}
\label{esdual}
\ee
The equations (\ref{sumas}) and (\ref{esdual}) give the useful relations:
$$
\tau_{0i}=i(a_{D,i}-\sum_j a^j\tau_{ji}), \,\,\,\,\,\ 
{\partial \tau_{0i} \over \partial a^k}=
-i\sum_j a^j {\partial \tau_{ij} \over 
\partial a^k},
$$
\be
{\partial \tau_{00} \over \partial a^k}=i\tau_{0k}-\sum_{ij}a^i a^j 
{\partial \tau_{ij} \over
\partial a^k}.
\label{rela}
\ee
Using now (\ref{mato}) one can prove that $s_D$ is a monodromy invariant,
\be
{\partial {\cal F} \over \partial s}=
i\Bigl(2 {\cal F} -\sum_i a^i a_{D,i}\Bigr)=
8 \pi b_1 u
\label{espurion}
\ee
and from (\ref{rela}) and (\ref{espurion}) we get
$$
\tau_{0i}=8 \pi b_1 {\partial u \over \partial a^i},
$$
\be
\tau_{00}=8 \pi i b_1 \Bigl(2u-\sum_i a^i{\partial u \over \partial a^i}\Bigr)
\label{taus}
\ee
Now we will present the transformation 
rules of the gauge couplings $\tau_{ij}$ under 
a monodromyy matrix $\Gamma$ in $Sp(2r, {\bf Z})$. In terms of 
the local coordinates $a_{\Gamma}^i=C^{ip}a_{D,p}(a^j,s)+ D^i_qa^q$ 
we have the couplings
\be
\tau_{\alpha \beta}^{\Gamma}=
{\partial^2 {\cal F} \over \partial a^{\alpha}_{\Gamma} 
\partial a^{\beta}_{\Gamma} }.
\label{dualcoup}
\ee
The change of coordinates is given by the matrix:
\be
\left(
      \begin{array}{cc} {\displaystyle {\partial a^i_{\Gamma} \over \partial
a^j}}& {\displaystyle {\partial
a^i_{\Gamma}
\over \partial s} }\\
& \\
{\displaystyle {\partial s \over  \partial a^j }} & 
{\displaystyle {\partial s
\over  \partial s}}
\end{array}\right) =
\left(\begin{array}{cc}{C^{ip} \tau_{pj}+D^i_j}&{C^{ip} \tau_{0p}}\\
                           0&1\end{array}\right),
\label{xv}
\ee   
with inverse
\be
\label{xvi}
\left( \begin{array}{cc} {\displaystyle{\partial a^i \over \partial
a^j_{\Gamma}}} & {\displaystyle {\partial
a^i \over \partial s }}  \\
 &  \\
     {\displaystyle {\partial s \over \partial a^j_{\Gamma}}} &
{\displaystyle {\partial s \over \partial s}}
\end{array}\right) =
 \left( \begin{array}{cc}\Big( (C\tau +D)^{-1} \Big)^i_j&
-\Big( (C\tau +D)^{-1} \Big)^i_kC^{kp}\tau_{p0}\\
                           0&1 \end{array}\right).
\ee 
Therefore we have:
$$
\Bigl( {\partial \over \partial a^j_{\Gamma}} \Big)_{\Gamma -{\rm
basis}}
=\Big( (C\tau +D)^{-1} \Big)^i_j{\partial \over \partial a^i},
$$
\be
\Big( {\partial \over \partial s} \Big)_{\Gamma -{\rm basis}}
={\partial \over \partial s}-\Big[(C\tau +D)^{-1}C \tau \Big]^i_0
{\partial \over \partial a^i};
\label{xvii}
\ee
which lead to the transformation rules for the couplings:
\bea
\tau_{ij}^{\Gamma}&=&\Big( A \tau + B \Big)\Big( C
\tau +D \Big)^{-1}_{ij},
 \,\,\,\,\,\,\,\,\,\ \tau_{0i}^{\Gamma}=
\tau_{0j}\Big( (C\tau +D)^{-1} \Big)^j_i,
\nonumber \\
\tau_{00} ^{\Gamma}&=&\tau_{00}- 
\tau_{0i}\Big[(C\tau +D)^{-1}C \tau \Big]^i_0.
\label{xviii}
\eea  
\section{Effective potential and vacuum structure}
\setcounter{equation}{0}  
In this section we will obtain, starting from the formalism 
developed in the previous section, the effective 
potential in the Coulomb phase of the softly 
broken $N=2$ theory, for a general group of 
rank $r$. 

To break $N=2$ down to $N=0$ we freeze 
the spurion superfield to a constant. 
The lowest component is 
fixed by the scale $\Lambda$, and we only 
turn on the auxiliary $F^0$ ({\it i.e.} we take 
$D^0=0$). We must include in the effective 
lagrangian $r+1$ vector multiplets, where $r$ 
is the rank of the gauge group:
\be
A^{\alpha}=(A^0, A^{I}), \,\,\,\,\,\,\ I=1, \cdots, r.
\label{twoi}
\ee
There are submanifolds in the moduli space where extra states become 
massless and we must include them in the effective lagrangian. They are 
BPS states corresponding to 
monopoles or dyons, so we introduce 
$n_H$ hypermultiplets near these submanifolds 
in the low energy description:
\be
(M_i, {\widetilde M}_i), \,\,\,\,\,\,\,\,\ i=1, \cdots, n_H
\label{twoii}
\ee
We suppose that these BPS states are mutually local, 
hence we can find a symplectic 
transformation such that they have $U(1)^r$ charges 
$(q_i^I,-q_i^I)$ with respect to the $I$-th $U(1)$ (we follow the 
$N=1$ notation). The full $N=2$ effective lagrangian contains two terms:
\be
{\cal L}={\cal L}_{\rm VM}+{\cal L}_{\rm HM}, 
\label{twoiii}
\ee
where ${\cal L}_{\rm VM}$ is given in (\ref{lagr}), and
\bea
{\cal L}_{\rm HM}&=&\sum_i\int d^4 \theta 
\big( M^{*}_{i}{\rm e}^{2q_i^IV^{(I)}}M_i 
 +{\widetilde M}^{*}_i{\rm e}^{-2q_i^IV^{(I)}}
{\widetilde M}_i\big)\nonumber\\
&+& \sum_{I,i}\Bigl( \int
d^2 \theta {\sqrt 2}A^{I}q_i^I M_i {\widetilde M}_i + {\rm h.c.} \Big)
\label{twoiv}
\eea      
The terms in (\ref{twoiii}) contributing to the effective potential are
\bea
V &=& b_{IJ}F^I {\overline F}^J + b_{0I} 
\big( F^0{\overline F}^I+{\overline F}^0F^I \big)+ 
b_{00}|F^0|^2 \nonumber\\
&+& {1 \over 2}b_{IJ} D^I D^J + 
D^I q_i^I (|m_i|^2-|{\widetilde m}_i|^2)+
 |F_{m_i}|^2+ |F_{{\widetilde m}_i}|^2 \nonumber \\
&+& \sqrt{2} \Bigl(F^I q_i^I m_i {\widetilde m}_i+
 a^Iq_i^I m_i F_{{\widetilde m}_i}+
a^Iq_i^I {\widetilde m}_i F_{m_i} +{\rm h.c.}\Bigr),
\label{twov}
\eea
where all repeated indices are summed 
and $b_{\alpha \beta}={\rm Im} \tau_{\alpha \beta} /4\pi$. 
We eliminate the auxiliary fields and obtain:
$$
D^I=-(b^{-1})^{IJ}q_i^J (|m_i|^2-|{\widetilde m}_i|^2),
$$
$$
F^I=-(b^{-1})^{IJ}b_{0J}F^0-
\sqrt{2}(b^{-1})^{IJ}q_i^J{\overline m}_i{\overline {\widetilde m}}_i,
$$
\be
F_{m_i}=
-\sqrt{2}{\overline a}^Iq_i^I{\overline {\widetilde m}}_i,
\,\,\,\,\,\,\,\ 
F_{{\overline m}_i}=-\sqrt{2}{\overline a}^Iq_i^I{\overline m}_i.
\label{aux}
\ee
We denote $(q_i,q_j)=\sum_{IJ}q_i^I (b^{-1})^{IJ}q_j^I$, 
$(q_i, b_0)=\sum_{IJ}q_i^I (b^{-1})^{IJ}b_{0J}$, 
$a \cdot q_i=\sum_I a^Iq_i^I$. 
Substituting in (\ref{twov}) we obtain:
\bea
V&=&{1 \over 2} \sum_{ij}(q_i,q_j)(|m_i|^2-|{\widetilde m}_i|^2)
(|m_j|^2-|{\widetilde m}_j|^2)+
2\sum_{ij}(q_i,q_j)m_i{\widetilde m}_i{\overline m}_j
{\overline {\widetilde m}}_j\nonumber\\
&+&2\sum_i|a \cdot q_i|^2(|m_i|^2+|{\widetilde m}_i|^2) +
\sqrt{2}\sum_i(q_i, b_0)\Big(F^0 m_i {\widetilde m}_i + 
{\overline F}^0{\overline m}_i{\overline {\widetilde m}}_i \Big)
\nonumber \\
&-&|F^0|^2{{\rm det}b_{\alpha \beta} \over {\rm det} b_{IJ}},
\label{twovi}
\eea
where ${\rm det} b_{\alpha \beta}/ {\rm det} b_{IJ}
=b_{00}-b_{0I}(b^{-1})^{IJ}b_{0J}$ is 
the cosmological term. This term in the 
potential is a monodromy invariant. To prove this it is 
sufficient to prove invariance under the generators 
of the symplectic group 
$Sp(2r, {\bf Z})$:
$$
\left(\matrix{A& 0 \cr
               0&(A^{\rm T})^{-1}\cr}\right), 
\,\,\,\,\,\,\,\ A\in Gl(r, {\bf Z}),
$$
\be
T_{\theta}=\left(\matrix{{\bf 1}& \theta \cr
               0&{\bf 1}\cr}\right),\,\,\,\,\,\,\,\ \theta_{ij} \in {\bf Z},
\,\,\,\,\,\,\,\,\ \Omega=\left(\matrix{0& {\bf 1} \cr
               -{\bf 1}&0\cr}\right).
\label{gener}
\ee
Invariance under $T_{\theta}$ and the matrix 
involving only $A$ is obvious, and for 
$\Omega$ one can check it easily.

The vacuum structure is determined by the minima of 
(\ref{twovi}). As in \cite{soft}, 
we first minimize with respect to $m_i$, ${\overline m}_i$:
\bea
{\partial V \over \partial{\overline m}_i}&=&
\sum_j(q_i,q_j)(|m_j|^2-|{\widetilde m}_j|^2)m_i +
2|a \cdot q_i|^2 m_i \nonumber\\
&+& 2\sum_{j}(q_i,q_j)m_j{\widetilde m}_j{\overline {\widetilde m}}_i+
\sqrt{2}{\overline F}^0 (q_i,b_0){\overline {\widetilde m}}_i=0,
\label{twovii}
\eea
\bea
{\partial V \over \partial{\widetilde {\overline m}_i} }&=&
\sum_j(q_i,q_j)(-|m_j|^2+|{\widetilde m}_j|^2){\widetilde m}_i +
2|a \cdot  q_i|^2{\widetilde m} _i \nonumber\\
&+& 2\sum_{j}(q_i,q_j)m_j{\widetilde m}_j{\overline m}_i+
\sqrt{2}{\overline F}^0 (q_i,b_0){\overline m}_i=0.
\label{twoviii}
\eea
Multiplying (\ref{twovii}) by ${\overline m}_i$, (\ref{twoviii}) by 
${\overline {\widetilde m}}_i$ and substracting, we get
\be
\sum_j(q_i,q_j)(|m_j|^2-|{\widetilde m}_j|^2)(|m_i|^2+|{\widetilde m}_i|^2) 
+
2|a \cdot q_i|^2 (|m_i|^2-|{\widetilde m}_i|^2)=0.
\label{twoix}
\ee
Suppose now that, for some indices 
$i \in I$, $|m_i|^2+|{\widetilde m}_i|^2 >0$. 
Multiplying (\ref{twoix}) by $|m_i|^2-|{\widetilde m}_i|^2$ 
and summing over $i$ we obtain
\be
\sum_{ij}(q_i,q_j)(|m_i|^2-|{\widetilde m}_i|^2)
(|m_j|^2-|{\widetilde m}_j|^2)=
-\sum_{i \in I}{2|a \cdot  q_i|^2 \over |m_i|^2+|{\widetilde m}_i|^2}
(|m_i|^2-|{\widetilde m}_i|^2)^2.
\label{twox}
\ee
The matrix $(b^{-1})^{IJ}$ is positive definite, 
and if the charge vectors $q_i^I$ are 
linearly independent it 
follows that the matrix $(q_i, q_j)$ is positive 
definite too. Then the l.h.s. of 
(\ref{twox}) is $\ge 0$ while the r.h.s. is $\le 0$. 
The only way for this equation to be 
consistent is if
\be
|m_i|=|{\widetilde m}_i|, \,\,\,\,\,\,\ i=1, \cdots, n_H.
\label{twoxi}
\ee
In this case we can write the equation (\ref{twovii}), after 
absorbing the phase of $F^0=f_0{\rm e}^{i\gamma}$ in ${\widetilde m}_i$, as:
\be
2|a \cdot q_i|^2 m_i + 
2\sum_{j}(q_i,q_j)m_j{\widetilde m}_j{\overline {\widetilde m}}_i+
\sqrt{2}f_0 (q_i,b_0){\overline {\widetilde m}}_i=0.
\label{twoxii}
\ee
Multiplying by ${\overline m}_i$ and summing over $i$, we obtain
\be
2\sum_i |a \cdot q_i|^2 |m_i|^2 +
\sqrt{2}f_0 \sum_i(q_i,b_0){\overline m}_i{\overline {\widetilde m}}_i=
-2\sum_{ij}(q_i,q_j)m_j{\overline m}_i{\widetilde m}_j
{\overline {\widetilde m}}_i,
\label{twoxiii}
\ee
hence $\sqrt{2}f_0 
\sum_i(q_i,b_0){\overline m}_i{\overline {\widetilde m}}_i$ is real. We 
can insert \label{twoxiii} in (\ref{twovi}) and get 
the following expression for the effective potential:
\be
V=-f_0^2{{\rm det}b_{\alpha \beta} \over {\rm det} b_{IJ}}-
2\sum_{ij}(q_i,q_j)m_j{\overline m}_i{\widetilde m}_j
{\overline {\widetilde m}}_i.
\label{twoxiv}
\ee
If (\ref{twoxi}) holds, we can fix the gauge 
in the $U(1)^r$ factors and write
\be
m_i=\rho_i, \,\,\,\,\,\,\ {\widetilde m}_i=\rho_i {\rm e}^{i\phi_i}
\label{twoxv}
\ee
and (\ref{twoxii}) reads:
\be
\rho_i^2 \Bigl(|a\cdot q_i|^2+
\sum_{j}(q_i,q_j)\rho^2_j{\rm e}^{i(\phi_j-\phi_i)}+
{f_0 (q_i,b_0) \over \sqrt{2}}{\rm e}^{-i\phi_i}\Bigr)=0.
\label{twoxvi}
\ee
Apart form the trivial solution $\rho_i=0$, we have:
\be
|a\cdot q_i|^2+\sum_{j}(q_i,q_j)\rho^2_j{\rm e}^{i(\phi_j-\phi_i)}+
{f_0 (q_i,b_0) \over \sqrt{2}}{\rm e}^{-i\phi_i}=0
\label{twoxvii}
\ee
and we can have a monopole (or dyon) VEV in some 
regions of the moduli space. Notice that 
for groups of rank $r>1$ there is a coupling 
between the different $U(1)$ factors and one
 needs a numerical study of the equation 
above once the values of the charges $q_i^I$ are known. 
In addition, the moduli 
space is in that case very complicated and explicit
 solutions for the prepotential and
 gauge couplings of the $N=2$ theory are
 difficult to find. However we still can have 
some qualitative information in many cases
 under some mild assumptions, as we will see.
\section{Vacuum structure of the $SU(2)$ Yang-Mills theory}
\setcounter{equation}{0}
\subsection{The Seiberg-Witten Solution}
In \cite{swone} Seiberg and Witten obtained the structure of the quantum 
moduli space of the $N=2$ $SU(2)$ Yang-Mills theory and also the exact 
solution for the prepotential ${\cal F}$ including all the 
non-perturbative corrections. Some of the properties of this solution are:
  
i) The moduli space ${\cal M}_u$ is parametrized by 
$u=\langle {\rm Tr} \phi^2\rangle$ and can be understood as the complex 
$u$-plane. The 
$SU(2)$ symmetry is never restored, and the theory stays
in the Coulomb phase throughout the moduli space.

ii) ${\cal M}_u$ has a symmetry $u \rightarrow -u$
 (the non-anomalous subset of the $U(1)_R$ group),
and at the points $u=\Lambda^2$, $-\Lambda^2$
singularities in the holomorphic prepotential ${\cal F}$ develop.
Physically they correspond respectively to a massless
monopole and dyon with charges $(q_e,q_m)=(0,1)$, $(-1,1)$.
Hence near $u=\Lambda^2$, $-\Lambda^2$ the correct
effective action should include together
with the photon vector multiplet monopole or dyon hypermultiplets.

iii) The vector
$^{t}v=(a_D, a)$ defines a flat $SL_2({\bf Z})$
vector bundle over the moduli space ${\cal M}_u$.
Its properties are determined by the singularities and
the monodromies around them. Since
${\partial ^2 {\cal F} / \partial a^2}$ or
${\partial a_D/ \partial a}$ is the coupling constant,
these data are obtained from the $\beta$-function in
the three patches: large-$u$, the Higgs phase,
the monopole and the dyon regions.
{}From the BPS mass formula \cite{XIV,XV}
the mass of a BPS state of charge $(q_e,q_m)$
(with $q_e$, $q_m$ coprime for the charge to be stable) is:
\be
M={\sqrt 2}|q_e a+q_ma_D|.
\label{threevi}
\ee
If at some point $u_0$ in ${\cal M}_u$, $M(u_0)=0$, the monodromy
around this point is given by \cite{swone,swtwo,groups}
\be
 \left(\begin{array}{c}a_{D} \\
                    a \end{array}\right) \rightarrow M(q_e, q_m)
\left(\begin{array}{c}a_D\\
               a \end{array}\right),
\label{threevii}
\ee
\be
M(q_e, q_m)=\left(\begin{array}{cc} 1+2q_eq_m & 2q_e^2 \\
                           -2q_m^2&1-2q_eq_m \end{array} \right).
\label{threeviii}
\ee
Also for large $u$, ${\cal F}$ is dominated by the
perturbative one loop contribution, obtained from the
one loop $\beta$-function:
\be
{\cal F}_{\rm 1- loop}(a)={i \over 2\pi} a^2{\rm ln}{a^2 \over
\Lambda}
\label{threeix}
\ee
Hence we also have monodromy at infinity.
The three generators of the monodromy are therefore:
\be
M_{\infty}= \left(\begin{array}{cc}-1& 2\\
               0&-1 \end{array} \right),\,\,\,\
M_{\Lambda^2}=\left(\begin{array}{cc}1& 0\\
                                 -2&1 \end{array}\right),
 \,\,\,\ M_{-\Lambda^2}=\left(\begin{array}{cc}-1& 2\\
                                                                -2&3
\end{array}\right);
\label{threex}
\ee
and they satisfy:
\be
M_{\infty}=M_{\Lambda^2}M_{-\Lambda^2}.
\label{threexi}
\ee
These matrices generate the subgroup
$\Gamma_2 \subset SL_2{\bf Z}$ of $2 \times 2$
matrices congruent to the unit matrix modulo $2$.

We learn from (\ref{threevi})-(\ref{threeviii}) that in the Higgs,
monopole and dyon patches, the natural independent
variables to use are respectively $a^{(h)}=a$, $a^{(m)}=-a_D$,
$a^{(d)}=a_D-a$. Thus in each patch we have a
different prepotential:
\be
{\cal F}^{(h)}(a), \,\,\,\ {\cal F}^{(m)}(a^{(m)}), \,\,\,\ {\cal
F}^{(d)}(a^{(d)}).
\label{threexii}
\ee

iv) The explicit form of $a(u)$, $a_D(u)$ is given in
terms of the
 periods of a meromorphic differential of the second
kind on a genus
 one surface described by the equation:
\be
y^2=(x^2-\Lambda^4)(x-u),
\label{threexiii}
\ee
describing the double covering of the plane branched at
${\pm \Lambda^2}$, $u$, $\infty$.
We choose the cuts $\{-\Lambda^2, \Lambda^2\}$, $\{u, \infty\}$.
 The correctly normalized meromorphic 1-form is:
\be
\lambda=\Lambda{{\sqrt 2} \over 2\pi} {dx {\sqrt {x-u/\Lambda^2}}
 \over {\sqrt {x^2-1}}}.
\label{threexiv}
\ee
Then:
\be
a(u)= \Lambda{{\sqrt 2} \over \pi} \int_{-1}^{1}{dt {\sqrt
{u/\Lambda^2-t}} \over {\sqrt {1-t^2}}};
\label{threexv}
\ee
\be
a_D(u)= \Lambda{{\sqrt 2} \over \pi} \int_{1}^{u/\Lambda^2}{dt {\sqrt
{u/\Lambda^2-t}} \over {\sqrt {1-t^2}}}.
\label{threexvi}
\ee
Using the hypergeometric representation of the elliptic functions
\cite{XVI}:
$$
K(k)={\pi \over 2}F(1/2, 1/2, 1;k^2); \,\,\ K'(k)=K(k');
$$
\be
\label{threexvii}
E(k)={\pi \over 2}F(-1/2, 1/2, 1;k^2); \,\,\ E'(k)=E(k'), \,\,\
{k'}^2+k^2=1,
\ee
we obtain :
\be
k^2={2 \over 1+u/\Lambda^2}, \,\,\,\ {k'}^2={u-\Lambda^2 \over
u+\Lambda^2},
\label{threexviii}
\ee
\be
a(u)={4 \Lambda \over \pi k}E(k), \,\,\,\,\,\ a_D(u)={4 \Lambda \over
i \pi }{E'(k)-K'(k) \over k}.
\label{threexix}
\ee
Using the elliptic function identities:
\be
{dE \over dk}={E-K \over k}, \,\,\,\,\ {dK \over dk}={1 \over k
{k'}^2}(E-{k'}^2 K),
\label{threexx}
\ee
\be
{dE' \over dk}=-{k \over {k'}^2}(E'-K'), \,\,\,\,\ {dK' \over
dk}=-{1\over k {k'}^2}(E'-{k}^2K'),
\label{threexxi}
\ee
the coupling constant becomes:
\be
\tau_{11}={\partial a_{D} \over \partial a}={da_{D}/dk \over
da/dk}={iK' \over K},
\label{threexxii}
\ee
which is indeed the period matrix of the curve (\ref{threexiii}).

\subsection{Vacuum structure of the softly broken $SU(2)$ theory}

When we softly break the $N=2$ $SU(2)$ Yang-Mills theory we obtain an 
effective potential including the couplings $\tau_{01}$ and $\tau_{00}$. 
In the normalization of \cite{swone}, and with $b_1=1/4\pi^2$, the 
spurion-induced couplings are
\be
\tau_{01}={2 \over \pi}{\partial u \over \partial a}, \,\,\,\,\,\,\,\,\ 
\tau_{00}={2i \over \pi}\Big( 2u-a{\partial u \over \partial a}\Big).
\label{twocoup}
\ee
The monodromy transformations of the couplings 
(\ref{xviii}) have a simple expression in the $SU(2)$ case:
\bea
\tau_{11}^{\Gamma}&=&{\alpha \tau_{11}+ \beta \over \gamma
\tau_{11}+\delta},
 \,\,\,\,\,\,\,\,\,\ \tau_{01}^{\Gamma}={\tau_{01} \over \gamma
\tau_{11}+\delta },\nonumber \\
\tau_{00} ^{\Gamma}&=&\tau_{00}-{\gamma \tau_{01}^2 \over \gamma
\tau_{11}+\delta}.
\label{threexxiii}
\eea
From the exact Seiberg-Witten solution (\ref{threexv}), (\ref{threexvi}) 
and the 
previous equations we can compute the couplings $\tau_{ij}$
in the Higgs and monopole region.

i) Higgs region:

$$
a_D^{(h)}={4 \Lambda \over i\pi}{E'-K' \over k},
\,\,\,\,\ a^{(h)}={4 \Lambda \over \pi k}E(k),
$$
\be
\tau^{(h)}_{11}={iK' \over K},\,\,\,\ \tau^{(h)}_{01}=
{2\Lambda \over kK},\,\,\,\ \tau^{(h)}_{00}=
-{8i \Lambda^2\over \pi} \Big( {E-K \over k^2K} + {1 \over 2}\Big).
\label{threexxiv}
\ee

ii) Monopole region:

$$
a_D^{(m)}={4 \Lambda \over \pi k}E(k),
\,\,\,\,\ a^{(m)}=-{4 \Lambda \over i\pi}{E'-K' \over k},
$$
\be
\tau^{(m)}_{11}={iK \over K'},\,\,\,\
\tau^{(m)}_{01}={2i\Lambda \over kK'},\,\,\,\
\tau^{(m)}_{00}={8i \Lambda^2\over \pi}
\Big( {E'\over k^2K'} -{1 \over 2}\Big).
\label{threexxv}
\ee

In the analysis of the effective potential (\ref{twovi}) we must first 
minimize with respect to the monopole (or dyon) field. For $r=1$ the 
equation for the VEV (\ref{twoxvii}) is
\be
\rho^2+b_{11}|a|^2+{ b_{01}{\rm e}^{-i \phi} f_0 \over {\sqrt
2}}=0, 
\label{monovev}
\ee
and the last term must be real so ${\rm e}^{-i\phi}=\epsilon=\pm 1$. The 
charge is $q=1$ in the $SU(2)$ Yang-Mills theory, both in the monopole and 
in the dyon regions. Apart from the solution $\rho=0$ we can have
\be
\rho^2=-b_{11}|a|^2-{ b_{01}\epsilon f_0 \over {\sqrt
2}}>0.
\label{threexxvi}
\ee
 Note that
$b_{11}={1 \over 4\pi} {\rm Im}\,\ \tau_{11}$
is always positive, and therefore (\ref{threexxvi}) determines
 a region in the $u$-plane where the monopoles
acquire a VEV. Depending on the sign of $b_{01}$ we choose the sign of
$\epsilon$. In fact we can replace (\ref{threexxvi})
by:
\be
\rho^2=-b_{11}|a|^2+{1\over {\sqrt 2}}|b_{01}|f_0>0
\label{threexxvii}
\ee
and $f_0$ is always measured in units of $\Lambda$.
Thus for the numerical plots we set $\Lambda =1$.
From (\ref{twoxiv}) we get the effective potential:
\be
V=-{2 \over b_{11}} \rho^4-{{\rm det}b \over b_{11}}f_0^2
\label{threexxviii}
\ee
This is good news. It implies that the
 region where the monopoles acquire a VEV
 is energetically favored, and we have a
 first order phase transition to confinement.
 Depending on the sign of $b_{01}$,  $m$ and
$\widetilde m$ are either aligned or antialigned.
 The $SU(2)_R$ symmetry of $N=2$ supersymmetry
is broken by the explicit off-diagonal term
$b_{01}m {\widetilde m}/b_{11}$ in (\ref{twovi})
and by the VEV $\rho \not= 0$. 

Where $\rho^2 \rightarrow 0$, the potential maps smoothly onto the potential
for the Higgs region, 
\be
V^{(h)}=-{{\rm det}b^{(h)} \over b^{(h)}_{11}}f_0^2,
\label{threexxix}
\ee
where, we recall, ${{\rm det}b/b_{11}}$ 
is monodromy-invariant. In the monopole region, a 
nonzero monopole VEV is favoured, and the effective 
potential is given by (\ref{threexxviii}) and written in
terms of magnetic variables:
\be
V^{(m)}=-{2 \over b^{(m)}_{11}} \rho^4-{{\rm det}b^{(m)}
 \over b^{(m)}_{11}}f_0^2
\label{threexxx}
\ee
where $b^{(h)}$, $b^{(m)}$ are given in
(\ref{threexxiv}), (\ref{threexxv}). 

 In the Higgs region, the
 effective potential is given by (\ref{threexxix}) and
\begin{figure}
\centerline{
\hbox{\epsfxsize=5cm\epsfbox{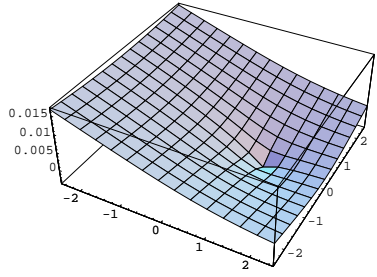}} }
\caption[]{Effective potential, $V^{(h)}$, (\ref{threexxix}).}
\figlabel\figone
\end{figure}
 \begin{figure}
\centerline{
\hbox{\epsfxsize=5cm\epsfbox{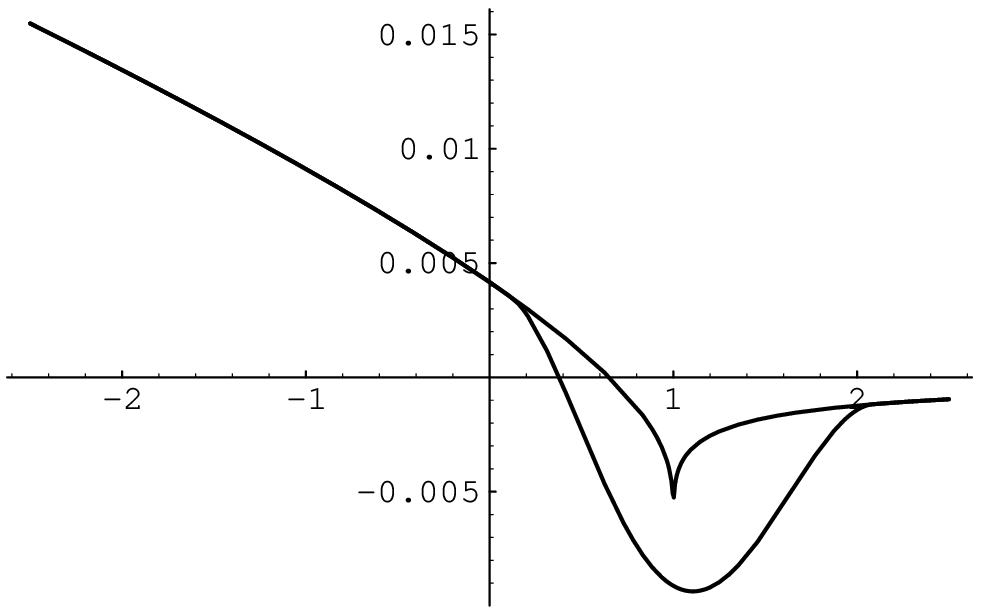}}\qquad
\hbox{\epsfxsize=5cm\epsfbox{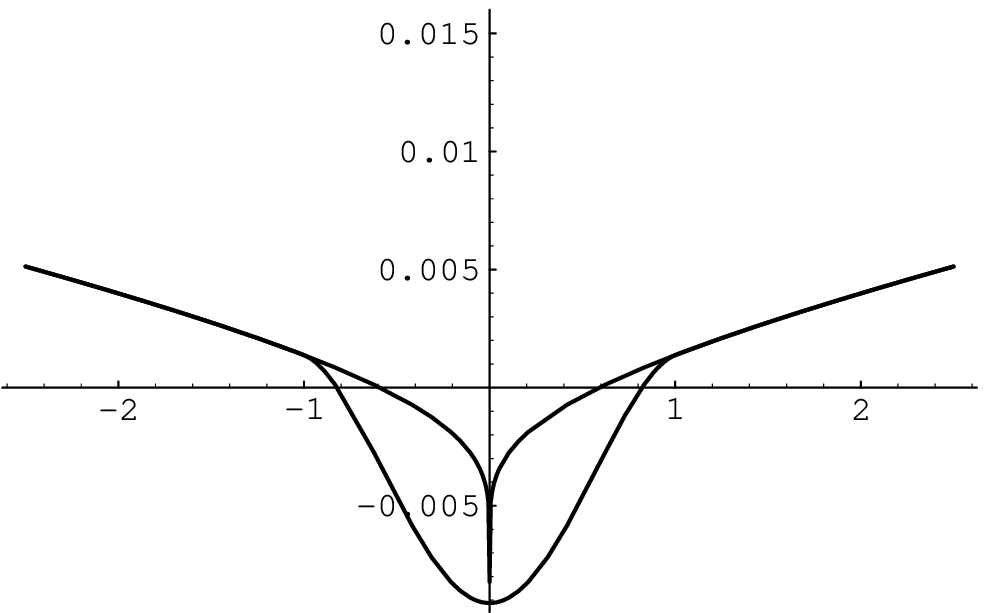}} }
\caption[]{Effective potential, $V^{(h)}$, (\ref{threexxix}) (top) and,
$V^{(m)}$, (\ref{threexxx}) (bottom) along
the real axis (left) and for $u=\Lambda^2(1+i y)$ (right). Both 
are plotted for
$f_0=0.3\Lambda$.}
\figlabel\figonea
\end{figure}
we plot it in \fig\figone. It has no minimum outside the monopole region near
$u=\Lambda^2$ (where, as we shall see, the energy can be further lowered by
giving the monopoles a VEV). One sees that the shape of
the potential makes the fields roll
towards the monopole region.
In \fig{\figonea}, we plot slices of the potential $V^{(h)}$ along the real
$u$-axis and parallel to the imaginary $u$-axis with ${\rm Re}(u)=\Lambda^2$.
For comparison, we also plot $V^{(m)}$. Note that they agree in the Higgs
region (where the monopole VEV vanishes), and that $V^{(m)}$ lowers the energy
(and smooths out the cusp in $V^{(h)}$ at $u=\Lambda^2$) in the monopole
region.

Next we look at the monopole
region (\ref{threexxvii}). $a$ ({\it i.e.} $a^{(m)}$) is a good
 coordinate in this region vanishing at $u=\Lambda^2$.
 As soon as $f_0$ is turned on monopole condensation
 \figalign{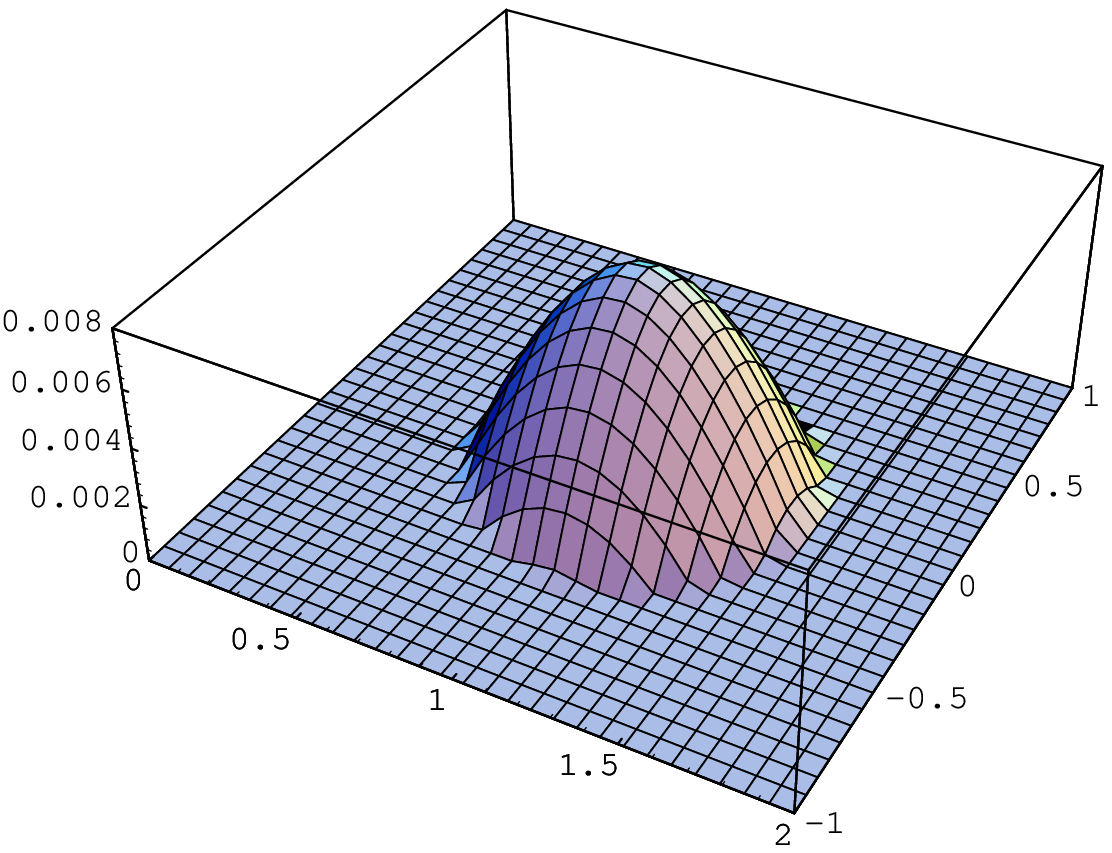}{\figtwo}{Monopole expectation value $\rho^2$ for
$f_0=0.1\Lambda$ on the $u$-plane.}{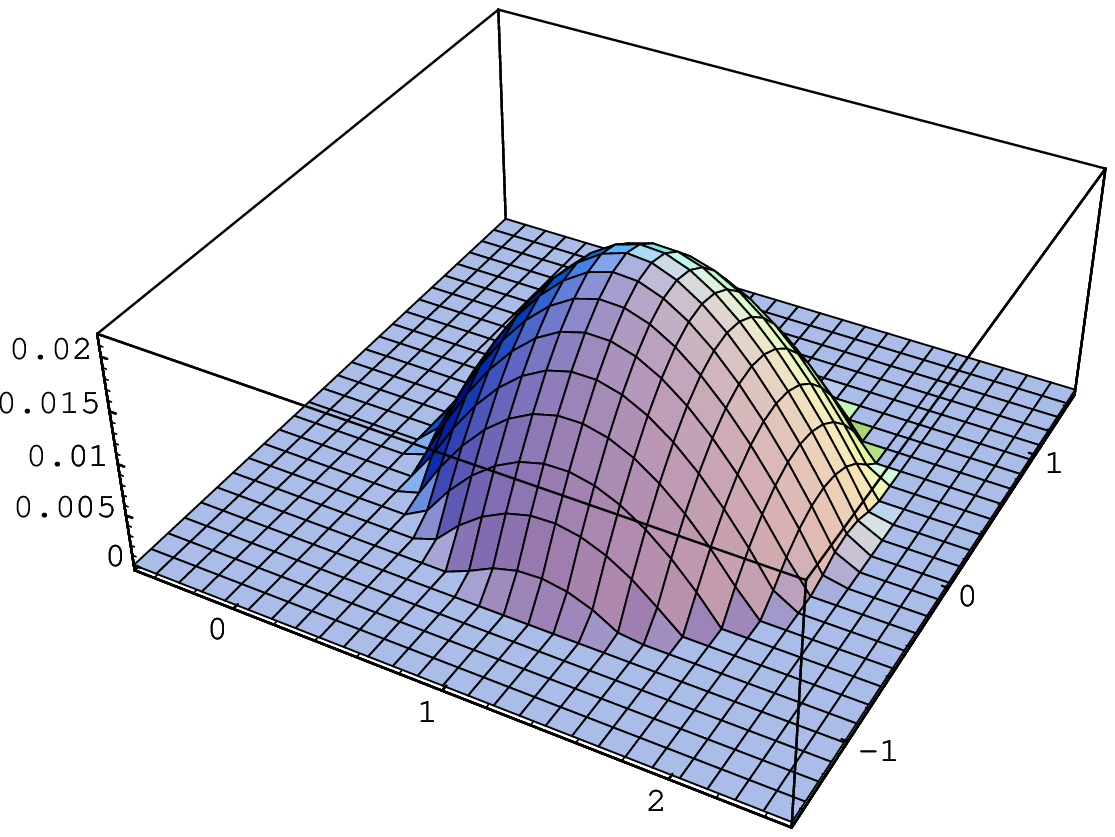}{\figthree}{Monopole
expectation value $\rho^2$ for $f_0=0.3\Lambda$ on the $u$-plane.}
and confinement occur. In \figs{\figtwo,\figthree}\ we
plot $\rho^2$ in the $u$-plane
for values of $f_0 = 0.1 \Lambda$,
$0.3 \Lambda$;
\figalign{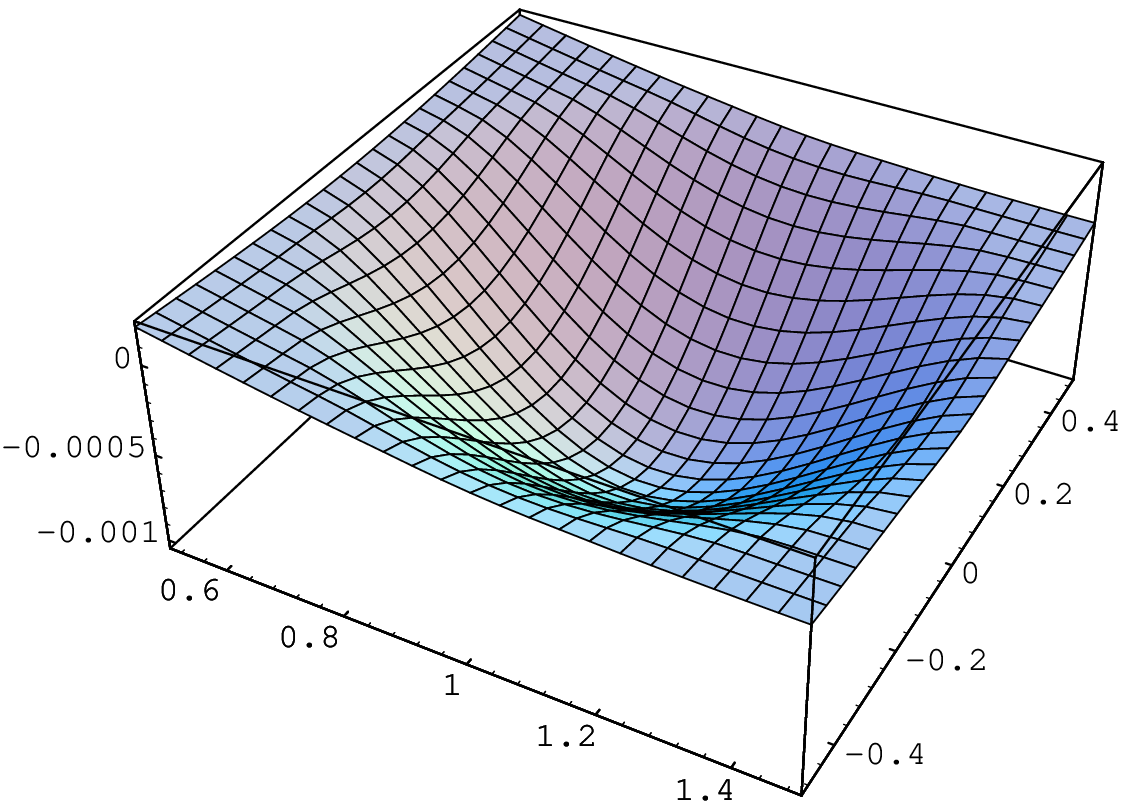}{\figfour}{Effective potential (\ref{threexxx}) for
$f_0=0.1
\Lambda$.}{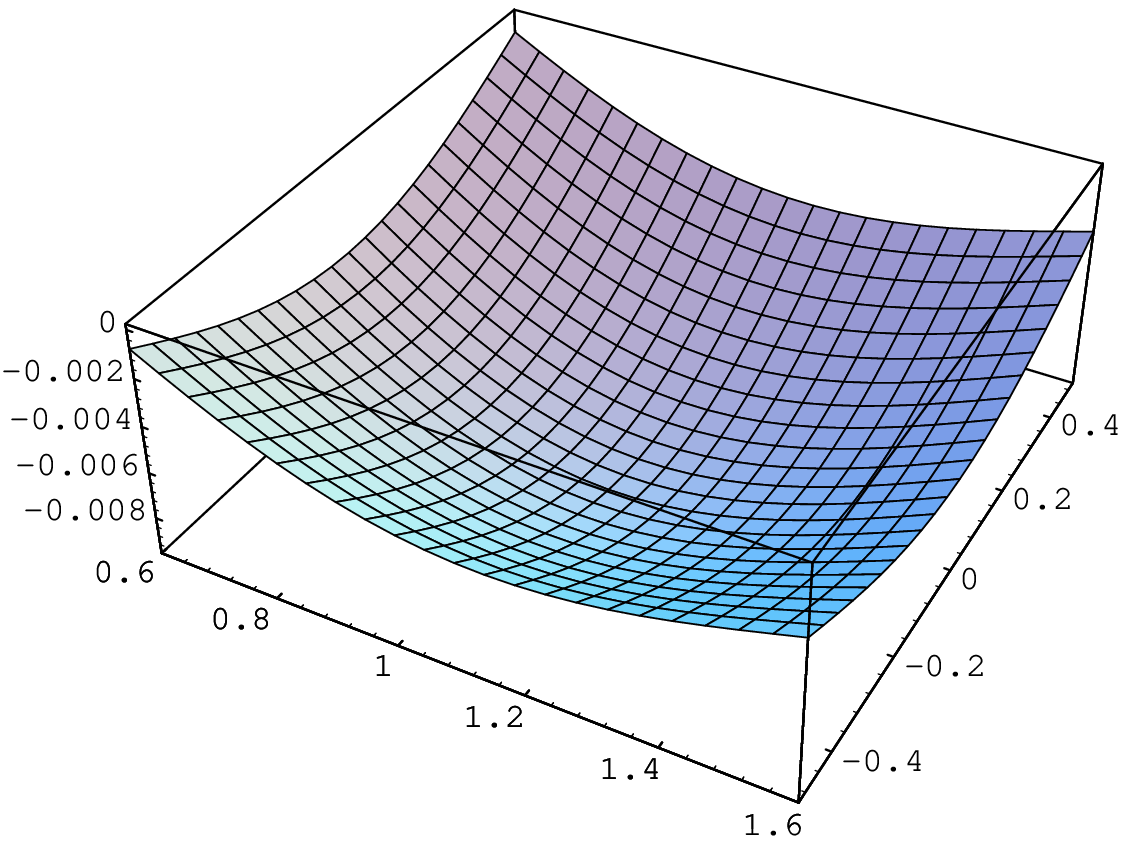}{\figfive}{Effective potential (\ref{threexxx}) for
$f_0=0.3 \Lambda$.}
and in \figs{\figfour,\figfive}\
the effective potential (\ref{threexxx}) for
the same values of the supersymmetry
 breaking parameter $f_0$.

One can see that the
minimum is stable and
that the size of the monopole VEV is $\sim f_0$.
There are two features
worth noticing. The first is that the absolute
minimum occurs along
the real $u$-axis. This is seen numerically
and also as
a consequence of the reality properties of
the elliptic functions. Second, as $f_0$ is increased, the region where
(\ref{threexxvii}) holds becomes wider.
\figalign{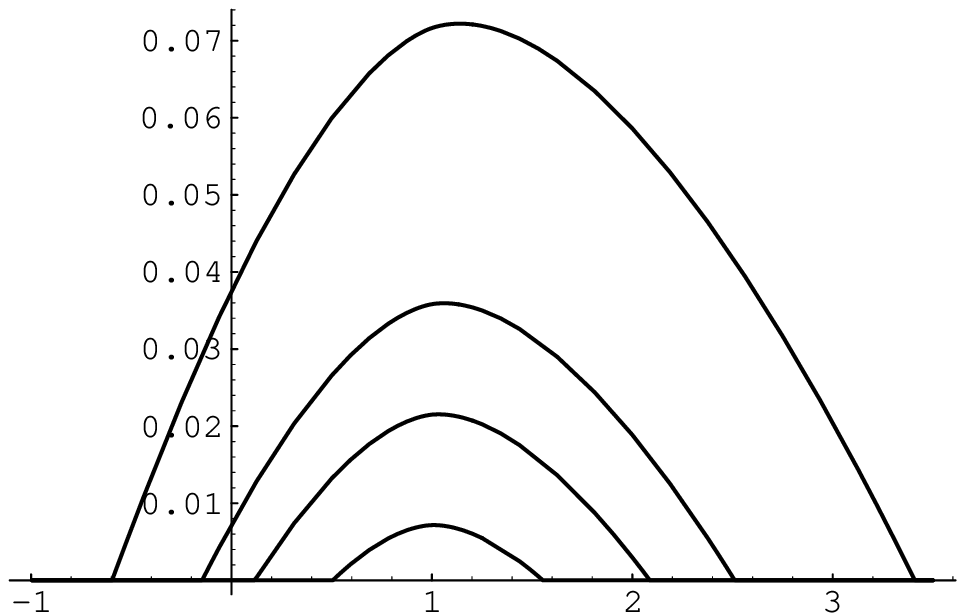}{\figsix}{Plot of $\rho^2$ along the real $u$-axis,
for $f_0/\Lambda=$ (from bottom to top) $0.1$, $0.3$, $0.5$, $1.0$.
}{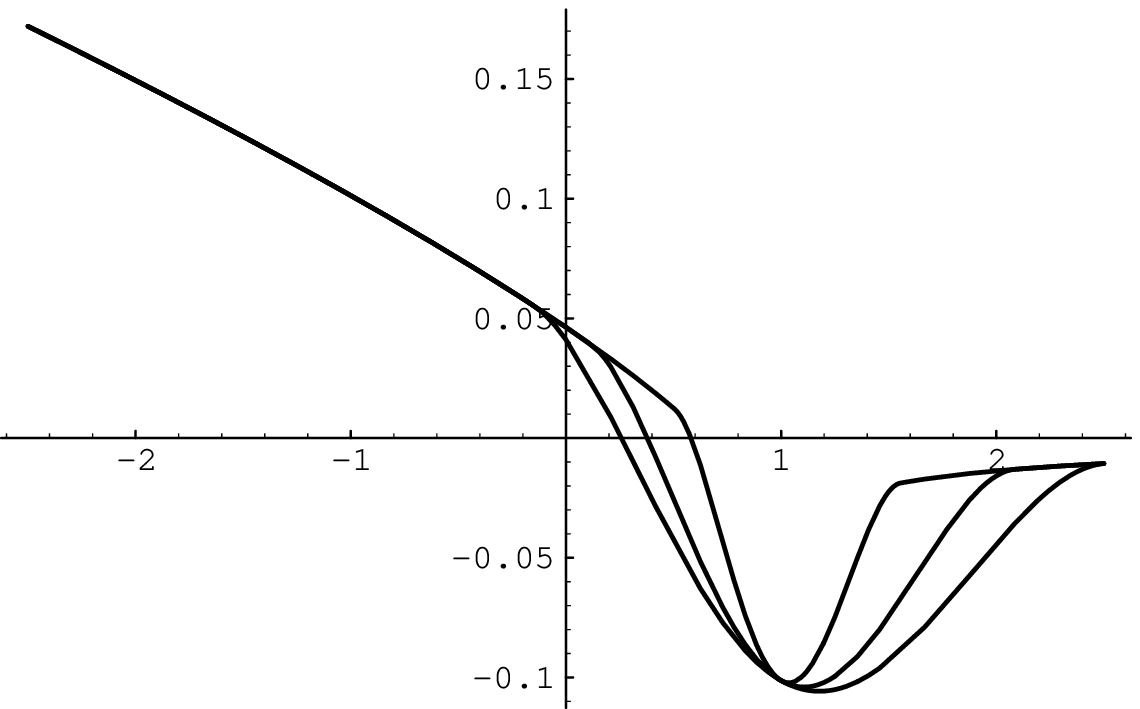}{\figseven}{$V^{(m)}/f_0^2$ along the real $u$-axis for
$f_0=0.1\Lambda$ (top), $0.5\Lambda$ (middle) and $\Lambda$ (bottom).}
 This is seen in \fig{\figsix}, where
$\rho^2$ is plotted along the real $u$-axis as a function of $f_0$.
Accordingly, the minimum of the effective potential moves to the right along
the real
$u$-axis, as one can see in \fig{\figseven}, where
$V^{(m)}/f_0^2$ is plotted for three increasing values of $f_0$ (we have
divided by $f_0^2$ to fit the three potentials on the same graph).

Finally, we turn to the dyon region. To understand what happens in the dyon
region, we study the transformation rules of the $\tau_{ij}$ couplings
under the residual ${\bf Z}_8 \subset U(1)_R$ symmetry whose generator
acts on the $u$-plane as $u \mapsto -u$.  The reason why we need
to analyze in general the behavior under ${\bf Z}_8$ is because
the representation we have chosen for the Seiberg-Witten
solution in sections 2,3 is  well adapted to study the
monopole region. Naively applying them to the dyon region, we may
encounter some discontinuities due to the position
of the cuts. Outside the curve of marginal stability
one can write the prepotential as \cite{swone}:
\be
{\cal F}={i\over 2\pi} a^2 \log{a^2\over \Lambda^2}+
a^2 \sum_{k\ge 1} c_k \Big( {\Lambda \over a}\Big)^{4k}.
\label{prepot}
\ee
If $\omega=e^{2\pi i/8}$ is the generator of the ${\bf Z}_8$
symmetry, it is easy to show that the couplings $\tau_{ij}$
transform according to\footnote{There is one more aspect of the ${\bf Z}_8$
transformation
rules worth noticing. If we implement these rules
we find that the condensate moves to the dyon region,
and one might be tempted to conclude that with this
choice it is the dyon that condenses.  This is not
the case. Using the one-loop $\beta$-function, we know
that $\Lambda^4 \sim {\rm exp}(-{8\pi^2\over g^2}+i\theta)$.
The action of ${\bf Z}_8$ amounts to the change
$\Lambda \mapsto i\Lambda$ or what is the same,
$\theta \mapsto \theta + 2\pi$.  Using the relation
found in \cite{dyon}, when we make this change the massless
state at $u=-\Lambda^2$ (before supersymmetry breaking)
has zero electric charge, while the state at $u=\Lambda^2$
acquires charge one. Thus we find again a monopole
condensate, in a way consistent with the ${\bf Z}_2$-symmetry.
}:
$$
 a\mapsto i a, \qquad a_D \mapsto i (a_D-a),
$$
\be
\tau_{11} \mapsto \tau_{11}-1, \qquad \tau_{01}\mapsto i \tau_{01}, \qquad
\tau_{00}\mapsto -\tau_{00}.
\label{symmetry}
\ee
So the relation between the dyon and monopole variables is:
\bea
a^{(d)}(u)=i a^{(m)}(-u),&\quad
a_D^{(d)}(u)=i \left(a_D^{(m)}(-u)-a^{(m)}(-u)\right),\label{dymon}\\
\tau_{11}^{(d)}(u)=\tau_{11}^{(m)}(-u)-1,&\quad\tau_{01}^{(d)}(u)=i
\tau_{01}^{(m)}(-u),\quad\tau_{00}^{(d)}(u)=-\tau_{00}^{(m)}(-u),\nonumber
\eea
 with $a_D^{(d)}=-a_D$. Using the expressions for
 the monopole couplings in (\ref{threexxv}), which are
well-behaved near $u=\Lambda^2$, we obtain 
expressions for the dyon couplings
which are well-behaved near $u=-\Lambda^2$.
The analysis of (\ref{threexxvii}) changes
crucially once these rules are implemented. Near the
monopole region $a^{(m)}\sim i(u-\Lambda^2)$, hence $\tau^{(m)}_{01}\sim i$
is purely imaginary.  In (\ref{threexxx}) although $b_{11}$ diverges
at $u=\Lambda^2$ the divergence is cancelled by the vanishing of
 $a^{(m)}$ at the same point.  Since ${\rm Im}\tau^{(m)}_{01} >0$ as soon
as $f_0\ne 0$ the monopoles condense.  Using (\ref{dymon}), however, we see
that $a^{(d)}\sim (u+\Lambda^2)$ with a real
coefficient.  Thus ${\rm Im}\tau_{01}^{(d)} =0$ at $u=-\Lambda^2$
and we conclude from (\ref{fiveix}) that the dyon condensate {\it vanishes}
along the real $u$-axis. Nevertheless, a dyon condensate {\it is}
energetically favoured in a pair of complex-conjugate regions in the $u$-plane
centered about $u=-\Lambda^2$.
\figalign{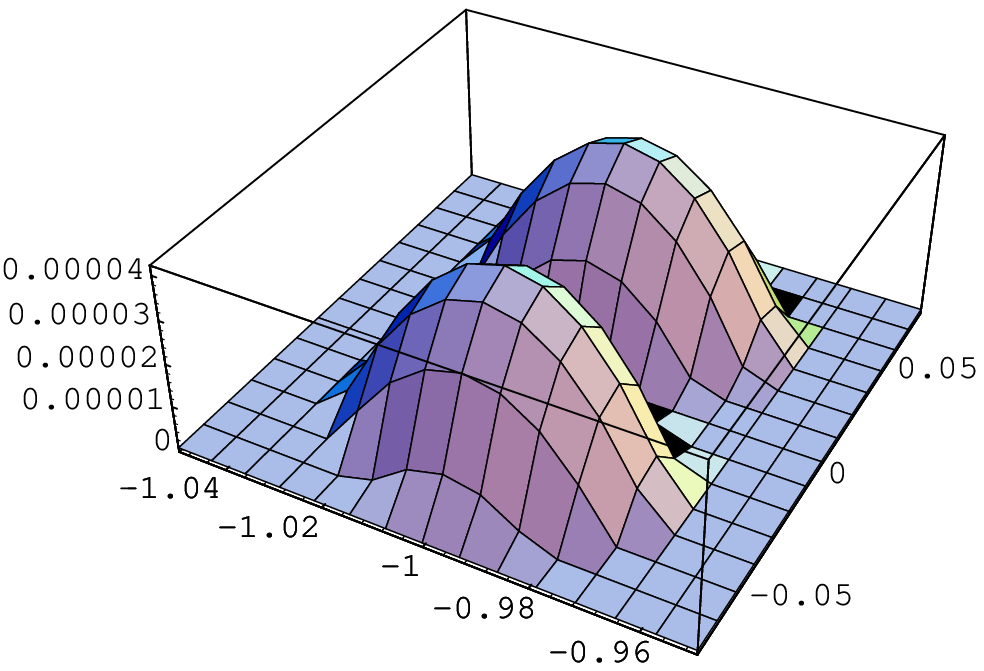}{\figeight}{Dyon expectation value $\rho_{(d)}^2$ for
$f_0=0.3\Lambda$ on the $u$-plane.}{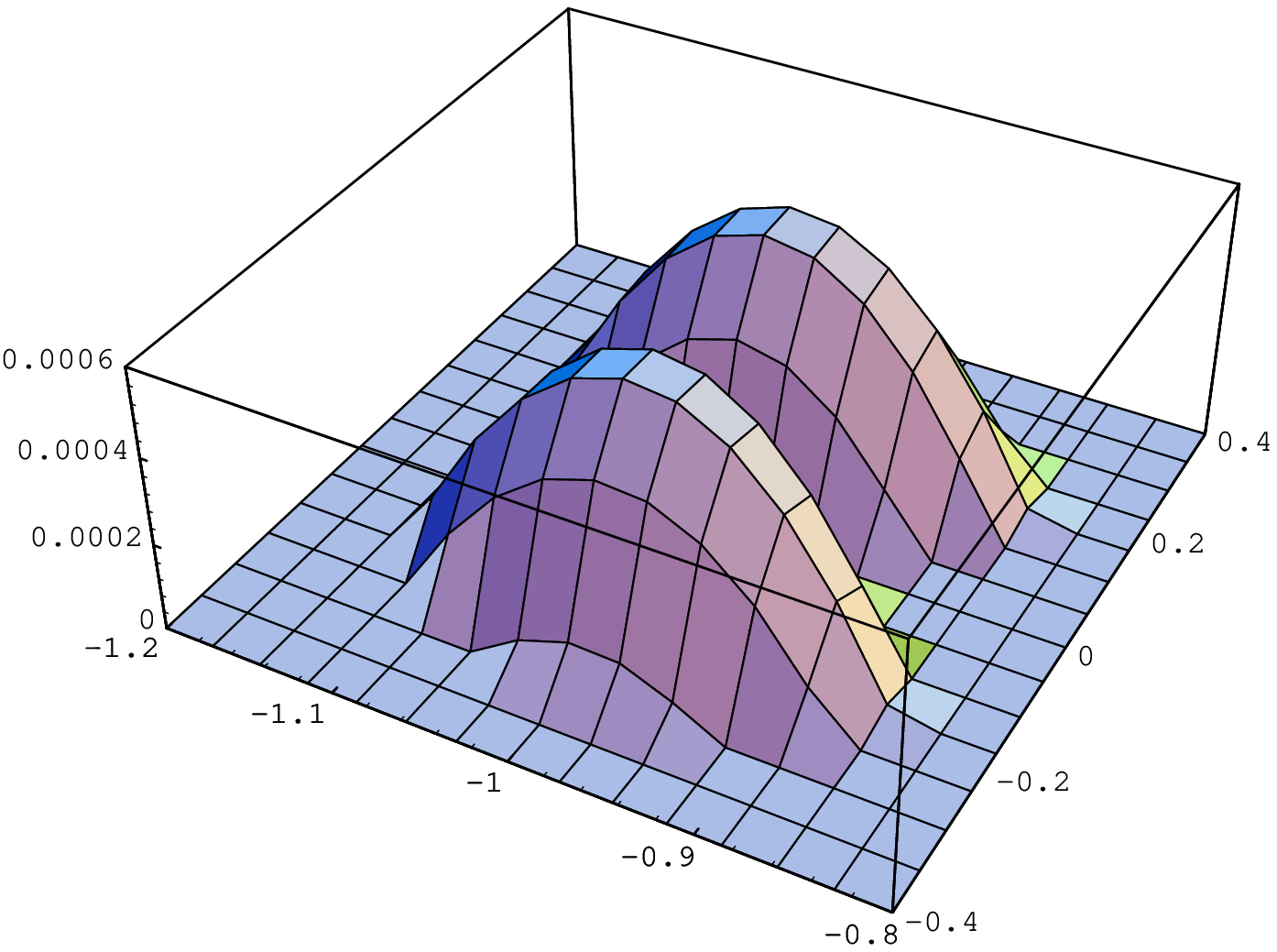}{\figeightb}{Dyon
expectation value $\rho_{(d)}^2$ for $f_0=\Lambda$ on the $u$-plane.}
We plot $\rho_{(d)}^2$, for two different values of $f_0$ in
\figs{\figeight,\figeightb}.

Unlike the monopole VEV, the magnitude of the dyon VEV is {\it tiny} on the
scale of $V^{(h)}$. It therefore makes an all-but-negligible contribution to
the
\begin{figure}
\epsfxsize=5cm
\centerline{\epsfbox{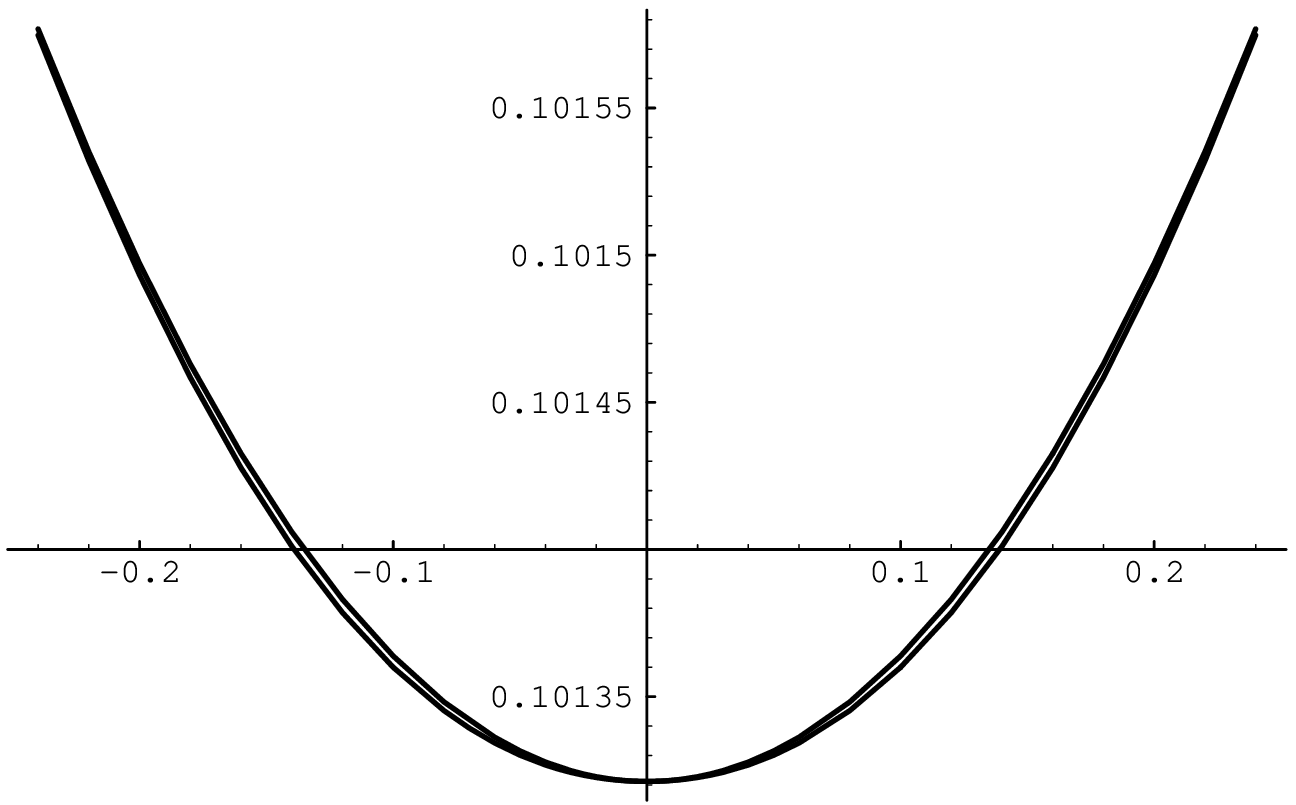} }
\caption[]{Plot of $V^{(h)}(u)$ (top) and $V^{(d)}(u)$ (bottom) versus ${\rm
Im}(u)$ for ${\rm Re}(u)=-\Lambda^2$ and $f_0=\Lambda$.}
\figlabel\figeightc
\end{figure}
effective potential (\fig\figeightc). In particular, $V^{(d)}$ does {\it not}
have a minimum in the dyon region. The only minimum of the full effective
potential is the one we previously found in the monopole region.

As we have already noted, the monopole region (in which
$\rho_{(m)}^2\neq0$) expands as $f_0$ is increased. Eventually, for $f_0\sim
1.3\Lambda$, it reaches the dyon region (in which
$\rho_{(d)}^2\neq0$). At this point, it is clear that our whole approximation
of including {\it just} the monopole field (or {\it just} the dyon field) in
the effective action breaks down.

What are the other limitations of our
approximations? First, we have neglected  certain soft supersymmetry breaking
terms which  arise when we derive the soft breaking terms 
from spontaneously broken $N=2$ supergravity. These 
additional terms scale to zero in the rigid limit, that is, they are
suppressed
by powers of ${\rm log}{\Lambda \over M_{\rm Pl}}$ or
$\Lambda \over M_{\rm Pl}$ and, for our purposes are negligible. We have also
neglected higher-spinor-derivative corrections to the Seiberg-Witten
effective action. These clearly cannot affect the vacuum structure
in the supersymmetric limit. They also,
{\it by definition} must be supersymmetric; otherwise they lead
to explicitly
hard supersymmetry breaking terms, which is
an entirely different matter from the soft
supersymmetry breaking we are considering. Nevertheless, once supersymmetry
is broken, they can, in principle, lead to corrections to the scalar
potential suppressed by higher powers of $f_0^2/\Lambda^2$. For the moderate
values of $f_0$ that we are considering, these corrections are numerically
rather small, and do not affect the qualitative features of the solutions
 we have found. {\it A priori}, if the higher spinor derivative terms in the
Seiberg-Witten effective action were known, we could systematically improve
our approximations by going to higher order in $f_0^2/\Lambda^2$.

However, the fundamental obstacle to pushing our approximation to larger values
of the soft supersymmetry breaking parameters would remain. The mutual
non-locality of the monopoles and dyons leads to our inability to
calculate the effective potential where the monopole and dyon regions overlap.
Since this is, at least initially, far from the monopole vacuum, we
expect that the monopole vacuum persists, at least as metastable minimum, even
beyond the critical value of $f_0$. But we do not know when (or if) a new,
lower minimum develops once the monopole and dyon regions overlap. If a new
vacuum does appear there, then we would have a first order phase transition to
this new confining phase 
\footnote{An explicit realization of this phase transition due to 
the overlapping of monopole and dyon regions 
occurs in the softly broken $SU(2)$ theory with 
one massless hypermultiplet \cite{agmm}}.
 This raises the  exciting possibility that
the correct description of the QCD vacuum requires the introduction of
mutually non-local monopoles and dyons. Phases of this nature have been shown
to arise in the
$N=2$
 moduli space for gauge group
$SU(3)$ \cite{ad}.
Perhaps the way to approach the true QCD vacuum in the
correct phase is to start with one of these $N=2$-superconformal field theories
and turn on a relevant, soft supersymmetry-breaking perturbation.

\section{Vacuum structure of the $SU(N)$ Yang-Mills theory}
\setcounter{equation}{0}
The moduli space of vacua of the $N=2$ $SU(N)$ Yang-Mills can be 
parametrized in a gauge-invariant way by the elementary symmetric polynomials 
$s_{l}$, $l=2, \cdots, N$ in the 
eigenvalues of $\langle \phi \rangle$, $\phi_i$. The vacuum 
structure of the theory is associated to 
the hyperelliptic curve \cite{kl}:
$$
y^2=P(x)^2-{\Lambda}^{2N},
$$
\be
P(x)={1 \over 2}{\rm det}(x-\langle \phi \rangle)=
{1 \over 2}\prod_i(x-\phi_i),
\label{fouri}
\ee
where $\Lambda$ is the dynamical scale of the $SU(N)$ theory and $P(x)$
 can be written in terms of the variables 
$s_l$ as $P(x)={1/2}\sum_l(-1)^ls_lx^{N-l}$. 
Once the hyperelliptic curve is known, one can 
compute in principle the metric 
on the moduli space and the exact quantum prepotential,
 but explicit solutions are 
difficult to find (they have been obtained in \cite{klt} for the $SU(3)$ case). 
But as in the $SU(2)$ case one expects that the minima
 of the effective potential for 
the $SU(N)$ theory are near the $N=1$ points 
(at least for a small supersymmetry 
breaking parameter). The physics of the $N=1$ points 
in $SU(N)$ theories has a much 
simpler description because it involves only small 
regions of the moduli space, and 
has been studied in \cite{ds}. The $N=1$ points correspond 
to points in the moduli 
space where $N-1$ monopoles coupling to each $U(1)$ 
become massless simultaneously. 
From the point of view of the hyperelliptic curve 
this corresponds to a 
simultaneous degeneration 
of the $N-1$ $\alpha$-cycles, associated to monopoles. This means in turn that 
the polynomial $P(x)^2-\Lambda^{2N}$ must have $N-1$ double zeros and two 
single zeros. If we set $\Lambda=1$, this can be achieved with the Chebyshev 
polynomials
\be
P(x)={\rm cos}\Big(N{\rm arccos} {x\over 2}\Big),
\label{fourii}
\ee
and the corresponding eigenvalues are 
$\phi_i =2 {\rm cos} \pi(i-{1 \over 2})/N$. 
The other 
$N-1$ points, corresponding to the simultaneous condensation of $N-1$ 
mutually local dyons, are obtained with the action of the anomaly-free 
discrete subgroup 
${\bf Z}_{4N} \subset U(1)_R$. One can perturb slightly the curve 
(\ref{fourii}) to obtain the 
effective lagrangian (or equivalently, 
the prepotential) at lowest order. What is found 
is that, in terms of the dual monopole variables 
$a_{D,I}$, the $U(1)$ factors are 
decoupled and $\tau^D_{IJ} \sim \delta_{IJ}\tau_I$. 
Near the $N=1$ point where $N-1$ 
monopoles 
become massless one can then simplify the equation 
(\ref{twoxvii}) for the monopole 
VEVs, because $q_i^I=\delta_i^I$, 
$(b^{-1})^{IJ}=\delta^{IJ} b^{-1}_I$. The equation reduces 
then to $r=N-1$ $SU(2)$-like equations, 
and in particular the phase factors 
${\rm e}^{-i\phi_I}$ must be real. We then 
set ${\rm e}^{-i\phi_I}=\epsilon_I$, 
$\epsilon_I=\pm 1$. The VEVs are determined by:
\be
 \rho^2_I=-b_I|a_{D,I}|^2-
{f_0 b_{0I}\epsilon_I \over \sqrt{2}},\,\,\,\,\,\,\ I=1, \cdots, r.
\label{fouriii}
\ee
The effective potential (\ref{twoxiv}) reads:
\be
V=-f_0^2 \Big(b_{00}-\sum_I{b_{0I}^2 \over b_I} \Big)
-2\sum_I{1 \over b_I}\rho_I^4.
\label{fouriv}
\ee
The quantities that control, at least qualitatively, 
the vacuum structure of the theory, 
are $b_{0I}$ and $b_{00}$. If $b_{0I} \not= 0$ 
at the $N=1$ points, we have a monopole VEV 
for $\rho_I$ around this point. If $b_{0I}= 0$, we still 
can have a VEV, as it happens in 
the $SU(2)$ case in the dyon region, but we expect that
it will be too tiny to produce a local minimum. When 
one has monopole condensation at one of these $N=1$ 
points in all the $U(1)$ factors, the 
value of the potential at this point is given by
\be
V=-f_0^2b_{00},
\label{fourv}
\ee
and if the local minimum is very near to the $N=1$ point,
 we can compare the energy of the 
different $N=1$ points according to (\ref{fourv}) 
and determine the true vacuum of the 
theory. Hence, to have a qualitative picture
 of the vacuum structure, and if we 
suppose that the minima of the effective potential
 will be located near the $N=1$ points, 
we only need to evaluate $b_{0I}$, $b_{00}$ at
 these points. This can be done using the 
explicit solution in \cite{ds} and the expressions (\ref{taus}). 

To obtain the correct normalization of the constant appearing in 
(\ref{espurion}) we can evaluate $\sum_{I}a_{D,I}da/du-ada_{D,I}/du$ 
in the $N=1$ points, obtaining the constant value $4 \pi i b_1$. 
 The value of the quadratic Casimir at the $N=1$ point described by 
(\ref{fourii}) is
\be
u=\langle {\rm Tr}\phi^2 \rangle =
4\sum_{i=1}^N {\rm cos}^2 {\pi(i-1/2) \over N} =2N,
\label{fourvi}
\ee
and the values at the other $N=1$ 
points are given by the action of ${\bf Z}_N$ ($u$ has 
charge $4$ under $U(1)_R$):
 $u^{(k)}=2\omega^{4k}N$, $\omega={\rm e}^{\pi i /2N}$ with 
$k=0, \cdots, N-1$. To compute $\tau_{0I}$ we must 
also compute $\partial u /\partial a_{D,I}$. 
Using the results of \cite{ds}, we have:
\be
{\partial u \over \partial a_{D,I}}=-4i {\rm sin}{\pi I \over N},
\label{fourvii}
\ee
and using $b_1=2N/16 \pi^2$, we obtain
\be
\tau_{0I}=4\pi b_1{\partial u \over \partial a_{DI}}=
 -{2Ni \over \pi}{\rm sin}{\pi I \over N}.
\label{fourviii}
\ee
At the $N=1$ point where $N-1$ monopoles condense, $a_{D,I}=0$, therefore
\be
\tau_{00}=8\pi i u ={2i \over \pi}N^2.
\label{fourix}
\ee
(\ref{fourviii}) indicates that 
monopoles condense at this point in all the 
$U(1)$ factors, but with different VEVs. 
This is a consequence the spontaneous 
breaking of the $S_N$ symmetry permuting the $U(1)$ factors \cite{ds}. 

To study the other $N=1$ points we must 
implement the ${\bf Z}_N$ symmetry in the 
$u$-plane. The local coordinates $a^{(k)}_I$
 vanishing at these points are given by 
a $Sp(2r, {\bf Z})$ transformation acting on 
the coordinates $a_I$, $a_{D,I}$ 
around the monopole point. The ${\bf Z}_N$ symmetry implies that
\be
{\partial u \over \partial a^{(k)}_{I}}(u^{(k)})=
\omega^{2k}{\partial u \over \partial a_{D,I}}(u^{(0)}),
\label{fourx}
\ee
and then we get
$$
b^{(k)}_{0I}={1 \over 4 \pi}{\rm Im}\tau^{(k)}_{0I}=
-{N \over 2\pi^2}{\rm cos}{\pi k \over N}{\rm sin}{\pi I \over N},
$$
\be
b^{(k)}_{00}=
{1 \over 4 \pi}{\rm Im}\tau^{(k)}_{00}=
{1 \over 2}\Bigl({N \over \pi} \Bigr)^2
{\rm cos}{2\pi k \over N}.
\label{fourxi}
\ee
The first equation tells us that generically we will have dyon 
condensation at all the $N=1$
points, and the second equation together with (\ref{fourv}) implies 
that the condensate 
of $N-1$ monopoles at $u=2N$ is energetically favoured, 
and then it will be the true 
vacuum of the theory. Notice that the ${\bf Z}_N$ 
symmetry works in 
such a way that the size of the condensate, given by 
$|{\rm cos}{\pi k \over N}|$, corresponds to an energy given by 
$-{\rm cos}{2\pi k \over N}$: as one should expect, the bigger the 
condensate the smaller its energy. In fact, for 
$N$ even the $N=1$ point corresponding to $k=N/2$ has 
no condensation. In this case the energy is still 
given by (\ref{fourv}), as the effective potential equals 
the cosmological term with $b_{0I}=0$, 
and is the biggest one.

\section{Mass formula in softly broken $N=2$ theories}
\subsection{A general mass formula}
In some cases the mass spectrum of a 
softly broken supersymmetric theory is such that the 
graded trace of the square of the mass matrix is zero as it 
happens in supersymmetric theories 
\cite{mass}. We will see in this section that 
this is also the case when we softly break 
$N=2$ supersymmetry with a dilaton spurion. 

We will then compute the trace of the squared mass 
matrix which arises from the effective lagrangian 
(\ref{twoiii}), once the supersymmetry breaking parameter is turned on. 
The fermionic content of the theory is 
as follows: we have fermions $\psi^I$, $\lambda^I$ 
coming from the $N=2$ vector multiplet $A^I$ 
(in $N=1$ language, $\psi^I$ comes from the $N=1$ 
chiral multiplet and $\lambda^I$ from the $N=1$ vector multiplet). 
We also have 
``monopolinos" $\psi_{m_i}$, $\psi_{{\widetilde m}_i}$ 
from the $n_H$ matter hypermultiplets. 
To obtain the fermion mass matrix, we just
 look for fermion bilinears in (\ref{twoiii}). From 
the gauge kinetic part and the K\"ahler potential in ${\cal L}_{\rm VM}$
 we obtain:
\be
{i \over 16 \pi} F^{\alpha} {\partial}_{\alpha}\tau_{IJ}\lambda^I \lambda^J +
{i \over 16 \pi} {\overline F}^{\alpha} 
{\partial}_{\alpha}\tau_{IJ}\psi^I \psi^J.
\label{fivei}
\ee
where $F^0=f_0$ and the auxiliary fields $F^I$ are given in (\ref{aux}).
 From the kinetic term and the superpotential in ${\cal L}_{\rm HM}$ we get:
\bea
& & i\sqrt{2}\sum_{i} q_i \cdot \lambda \bigl({\overline m}_i\psi_{m_i} - 
{\overline{\widetilde m}}_i \psi_{{\widetilde m}_i} \bigr) \nonumber\\
&-& \sqrt{2} \sum_i \Bigl(a \cdot q_i\psi_{m_i}\psi_{{\widetilde m}_i}+
q_i \cdot \psi \psi_{{\widetilde m}_i} m_i +
q_i \cdot \psi \psi_{m_i} {\widetilde m}_i \Bigr)
\label{fiveii}
\eea
If we order the fermions as $(\lambda, \psi, 
\psi_{m_i}, \psi_{{\widetilde m}_i})$ and
denote $\mu^{IJ}=i F^{\alpha} {\partial}_{\alpha}\tau_{IJ}/4\pi$, 
${\hat \mu}^{IJ}=i{\overline F}^{\alpha} {\partial}_{\alpha}\tau_{IJ}/4\pi$, 
the ``bare" fermionic mass matrix reads:
\be
M_{1/2}= \left( \begin{array}{cccc} \mu /2& 0 & i\sqrt{2} q_i^I{\overline m}_i
&-i\sqrt{2} q_i^I{\overline {\widetilde m}}_i \\
0& {\hat \mu}/2& -\sqrt{2} q_i^I{\widetilde m}_i& -\sqrt{2} q_i^I m_i\\
i\sqrt{2} q_i^I{\overline m}_i&-
\sqrt{2} q_i^I{\widetilde m}_i&0&-\sqrt{2}a \cdot q_i\\
-i\sqrt{2} q_i^I{\overline {\widetilde m}}_i&
-\sqrt{2} q_i^I m_i&-\sqrt{2}a \cdot q_i& 0 \end{array}\right),
\label{fiveiii}
\ee
but we must take into account the wave function renormalization 
for the fermions $\lambda^I$, 
$\psi^I$ and consider
\be
{\cal M}_{1/2} =Z M_{1/2} Z, \,\,\,\,\,\,\,\ 
Z=\left( \begin{array}{cccc} b^{-1/2} & 0 & 0
&0 \\
0& b^{-1/2}& 0&0\\
0&0&1&0\\
0&
0&0& 1 \end{array}\right)
\label{fiveiv}
\ee
The trace of the squared fermionic matrix can be easily computed:
\bea
{\rm Tr}{\cal M}_{1/2} {\cal M}^{\dagger}_{1/2}&=&
{1\over 4}{\rm Tr}[\mu b^{-1} {\overline \mu}b^{-1}+
{\hat \mu}b^{-1}{\overline {\hat \mu}} 
b^{-1}]\nonumber\\
&+& 4\sum_i |a \cdot q_i|^2 + 
8\sum_i(q_i, q_i)(|m_i|^2+|{\widetilde m}_i|^2).
\label{fivev}
\eea
The scalars in the model are the monopole fields 
$m_i$, ${\widetilde m}_i$ and 
the lowest components of the $N=1$ chiral 
superfields in the $A^I$, $a^I$. To compute the
 trace of the scalar mass matrix we need
$$
{\partial^2 V \over \partial m_i \partial {\overline m}_i}= 
\sum_l (q_i, q_l) 
(|m_l|^2-|{\widetilde m}_l|^2)+ 
(q_i, q_i)(|m_i|^2+2|{\widetilde m}_i|^2)+2|a \cdot q_i|^2,
$$
$$
{\partial^2 V \over \partial {\widetilde m}_i \partial
{\overline {\widetilde m}_i} }= -\sum_l (q_i, q_l) 
(|m_l|^2-|{\widetilde m}_l|^2)+ 
(q_i, q_i)(2|m_i|^2+|{\widetilde m}_i|^2)+2|a\cdot q_i|^2,
$$
\bea
{\partial^2 V \over \partial a^I \partial {\overline a}_J}
&=&f_0^2{\partial^2 (b_0,b_0) \over \partial a^I \partial {\overline a}_J}+
2\sum_{k,l}{\partial^2 (q_k,q_l) \over \partial a^I \partial {\overline a}_J} 
m_k{\widetilde m}_k {\overline m}_l {\overline {\widetilde m}}_l\nonumber\\
&+&2\sum_{k}q_k^Iq_k^J(|m_k|^2+|{\widetilde m}_k|^2)\nonumber\\
&+&
\sqrt{2}\sum_k{\partial^2 (q_k,b_0)  
\over \partial a^I \partial {\overline a}_J}f_0
(m_k{\widetilde m}_k +{\overline m}_k{\overline{\widetilde m}}_k).
\label{fivevi}
\eea
In the last expression we used that, due to the holomorphy of the 
couplings $\tau_{\alpha \beta}$, 
$\partial^2_{I {\overline J}}b_{\alpha \beta}=0$.
 If we assume that we are in the conditions of section 2, 
at the minimum we have $|m_i|=
|{\widetilde m}_i|$, and the trace of the squared scalar matrix is
\be
{\rm Tr}{\cal M}^2_{0}=
 6\sum_i (q_i,q_i)(|m_i|^2+|{\widetilde m}_i|^2)+8\sum_i |a\cdot q_i|^2+ 
2(b^{-1})^{IJ}{\partial^2 V \over \partial a^I \partial {\overline a}_J},
\label{fivevii}
\ee
where we have included the wave function renormalization for the 
scalars $a^I$. The mass of the dual photon is given by the monopole 
VEV through the magnetic Higgs 
mechanism:
\be
{\rm Tr}{\cal M}^2_1=2\sum_i(q_i, q_i) ( |m_i|^2+|{\widetilde m}_i|^2).
\label{fiveviii}
\ee
Taking into account all these contributions, 
the graded trace of the squared matrix is:
\bea
& &\sum_j (-1)^{2j}(2j+1){\rm Tr }{\cal M}^2_j=
-{1\over 2}{\rm Tr}[\mu b^{-1} {\overline \mu}b^{-1}
+{\hat \mu}b^{-1}{\overline {\hat \mu}} 
b^{-1}] \nonumber\\
& &\,\,\,\,\,\,\,\ +
2f_0^2{\rm Tr}b^{-1}\partial {\overline \partial}(b_0,b_0)+
4\sum_{k,l}{\rm Tr}b^{-1}\partial {\overline \partial}(q_k,q_l)
m_k{\widetilde m}_k {\overline m}_l {\overline {\widetilde m}}_l\nonumber\\
& &\,\,\,\,\,\,\,\ +
2\sqrt{2}\sum_k{\rm Tr}b^{-1}\partial {\overline \partial}(q_k,b_0) f_0
(m_k{\widetilde m}_k +{\overline m}_k{\overline{\widetilde m}}_k).
\label{traza}
\eea
To see that this is zero, we write the bilinears in the monopole 
fields in terms of the auxiliary fields 
$F^I$, ${\overline F}^I$, using (\ref{aux}):
\be
\sum_{i} q_i^I{\overline m}_i{\overline {\widetilde m}}_i=
-{1 \over \sqrt{2}}(b_{IJ}F^J+b_{0I}f_0).
\label{fiveix}
\ee
Then we can group the terms in (\ref{traza}) depending on the number of 
$F^I$, ${\overline F}^I$, and check 
that they cancel separately. For instance, for the 
terms with two auxiliaries, we have from the first term in (\ref{traza}):
\be
-2(F^I {\overline F}^J +{\overline F}^I F^J)\partial_I b_{MN}(b^{-1})^{NP}
\partial_{\overline J}b_{PQ}(b^{-1})^{QM}
\label{fivex}
\ee
and from the third term
\bea
& &2F^I {\overline F}^J\partial_M b_{JN}(b^{-1})^{NP}
\partial_{\overline Q}b_{PI}(b^{-1})^{QM}\nonumber\\
&+&2F^I {\overline F}^J\partial_M b_{PI}(b^{-1})^{NP}
\partial_{\overline Q}b_{JN}(b^{-1})^{QM}.
\label{fivexi}
\eea
Taking into account the holomorphy of the couplings and the K\"ahler 
geometry, we have 
$\partial_M b_{PI}=\partial_I b_{PM}$, 
$\partial_{\overline Q}b_{JN}=
\partial_{\overline J}b_{QN}$, so (\ref{fivex}) and 
(\ref{fivexi}) add up to zero. With a little more algebra one can verify 
that the terms with one $F^I$ (and their conjugates with ${\overline F}^I$)
 and without any auxiliaries add up to zero too. The result is then:
\be
\sum_j (-1)^{2j}(2j+1){\rm Tr }{\cal M}^2_j=0.
\label{fivexii}
\ee

\subsection{Mass spectrum in the $SU(2)$ case}
In the $SU(2)$ case we can obtain much more information about the mass matrix 
and also determine its eigenvalues. First we consider the fermion mass
matrix. Taking into account that at the minimum of the effective potential 
$m = {\overline m}=\rho$, ${\tilde m}=\epsilon m$, we can introduce the linear 
combination:
\be
\eta_{\pm}={1\over {\sqrt 2}} (\psi_m \pm \epsilon \psi_{\tilde m}).
\label{fivexiii}
\ee
With respect to the new fermion fields $(\lambda, \eta_{+}, \psi,\eta_{-})$, 
the bare fermion mass matrix reads:
\be
M_{1/2}= \left( \begin{array}{cccc} {1 \over 2} \mu & -2\epsilon \rho & 0
&0\\
-2\epsilon \rho&-{\sqrt 2}\epsilon a & 0& 0\\
0&0&{1 \over 2} \mu &2i\rho\\
0&0&2i\rho & -{\sqrt 2}\epsilon a  \end{array}\right),
\label{fivexiv}
\ee
Notice that, in the $SU(2)$ case, the auxiliary field 
$F$ is real and $\mu ={\hat \mu}$. ${\cal M}_{1/2} {\cal M}^{\dagger}_{1/2}$ 
can be 
easily diagonalized . From (\ref{fivexiv}) it is easy to see
that the squared fermion mass matrix is block-diagonal with the 
same $2\times 2$ matrix in both entries:
\be
\left(\begin{array}{cc} b_{11}^{-2}\mu {\overline \mu}/4 
+4b_{11}^{-1}\rho^2& -\epsilon b_{11}^{-3/2}\mu\rho+
 2{\sqrt 2}{\overline a}\rho \\
-\epsilon b_{11}^{-3/2}{\overline \mu}\rho+
 2{\sqrt 2} a\rho &4b_{11}^{-1}\rho^2+2|a|^2\end{array} \right).
\label{fermimat}
\ee
Hence there are two different eigenvalues doubly degenerated. In terms of 
the determinant and trace of (\ref{fermimat}),
\bea
\alpha &=& (m^{F}_1)^2+(m^{F}_2)^2=
{1\over 4b_{11}^2}\mu{\overline \mu}+
2|a|^2+{8\over b_{11}}\rho^2\nonumber\\
\beta &=&(m^{F}_1)^2(m^{F}_2)^2 ={1\over b_{11}^2}|4\rho^2 + {\epsilon \over 
{\sqrt 2}}a\mu|^2,
\label{fivexv}
\eea
the eigenvalues are:
\be
(m^{F}_{1,2})^2={\alpha \over 2}\pm {1 \over 2}{\sqrt {\alpha^2-4\beta}}.
\label{fivexvi}
\ee
The computation of the scalar mass matrix is more lengthy. First we must 
compute the second derivatives of the effective potential, evaluated at the 
minimum. To obtain more simple expressions, we can use the identities 
(\ref{taus}) to express all the derivatives of the couplings in terms 
only of $\partial b_{11}/\partial a$, $\partial^2 b_{11}/\partial a^2$.
The results are:
\bea
{\partial^2 V \over \partial m \partial {\overline m}}&=&{3\over b_{11}}\rho^2
+2|a|^2, \,\,\,\,\,\ {\partial^2 V/\partial m^2}=
{\partial^2 V \over \partial {\widetilde m}^2}={1\over b_{11}}\rho^2, 
\,\,\,\,\,\ \nonumber \\ 
{\partial^2 V \over \partial m \partial {\widetilde m}}&=&
{\epsilon\over b_{11}}\rho^2+ {{\sqrt 2} b_{01} \over b_{11}}f_0, 
\,\,\,\,\,\ 
{\partial^2 V\over \partial m \partial {\overline{\widetilde m}}}=
{\epsilon\over b_{11}}\rho^2 \nonumber\\  
{\partial^2 V \over \partial m \partial a}&=& 
2\rho\Big[{\overline a}-
\big(b_{11} {\partial \over \partial a}{1 \over b_{11}}
\big) \big(|a|^2-{i \epsilon \over {\sqrt 2}}af_0 \big)\Big]\nonumber\\
{\partial^2 V \over \partial a^2}&=&
-b_{11}^2 \big({\partial \over \partial a}{1 \over b_{11}}\big) f_0
(af_0+2 {\sqrt 2} i\epsilon |a|^2)\nonumber \\
& & -b_{11}^2 \big({\partial^2 \over \partial a^2}{1 \over b_{11}}\big)
(af_0+{\sqrt 2}i\epsilon |a|^2)^2,\nonumber\\
{\partial^2 V \over \partial {\widetilde m} \partial a}&=&
\epsilon {\partial^2 V \over \partial m \partial a},\,\,\,\,\,\ 
{\partial^2 V \over \partial {\overline m} \partial a}=
 {\partial^2 V \over \partial m \partial a},\,\,\,\,\,\
{\partial^2 V \over \partial {\overline {\widetilde m}} \partial a}=
{\partial^2 V \over \partial {\widetilde m} \partial a},\nonumber\\
{\partial^2 V \over \partial {\overline a} \partial a}&=&
4\rho^2+{1 \over 2b_{11}}\mu {\overline \mu},
\label{fivexvii}
\eea
and the rest of the derivatives are obtained through complex 
conjugation. In the last line 
we used the result of the previous section.
 To obtain the bosonic mass matrix we must take into account the wave-function 
renormalization of the $a$, ${\overline a}$ 
variables, as in (\ref{fivevii}).  
Its eigenvalues are as follows: we have a zero 
eigenvalue corresponding to 
the Goldstone boson of the spontaneously broken 
$U(1)$ symmetry. There is 
also an eigenvalue with degeneracy two given by:
\be
2\big({\partial^2 V \over \partial m \partial {\overline m}}-
{\partial^2 V \over \partial {\widetilde m}^2}\big)= 
-{2 {\sqrt 2} \epsilon \over b_{11}}f_0b_{01}.
\label{fivexviii}
\ee
Notice that this is always positive if we have a non-zero VEV for $\rho$. 
The other three eigenvalues are best 
obtained numerically, as they are the solutions 
to a  
third-degree algebraic equation. 

As an application of these general results, 
we can plot the mass spectrum 
as a function of the supersymmetry breaking 
parameter $f_0$ in the $SU(2)$ Yang-Mills case, 
where the minimum corresponds to the monopole region 
and $\epsilon=-1$. We have only to 
compute the derivatives of the magnetic coupling, with the result:
\be
{\partial \tau^{(m)}_{11} \over \partial a^{(m)}}={\pi^2 \over 8}{k \over 
k'^2 K'^3},\,\,\,\,\,\
{\partial^2 \tau^{(m)}_{11} \over \partial {a^{(m)}}^2}=
-{\pi i \over 32} {k^2 \over k'^4 K'^4}\Big( k'^2-k^2+{3E' \over K'}\Big).
\label{fivexix}
\ee

\figalign{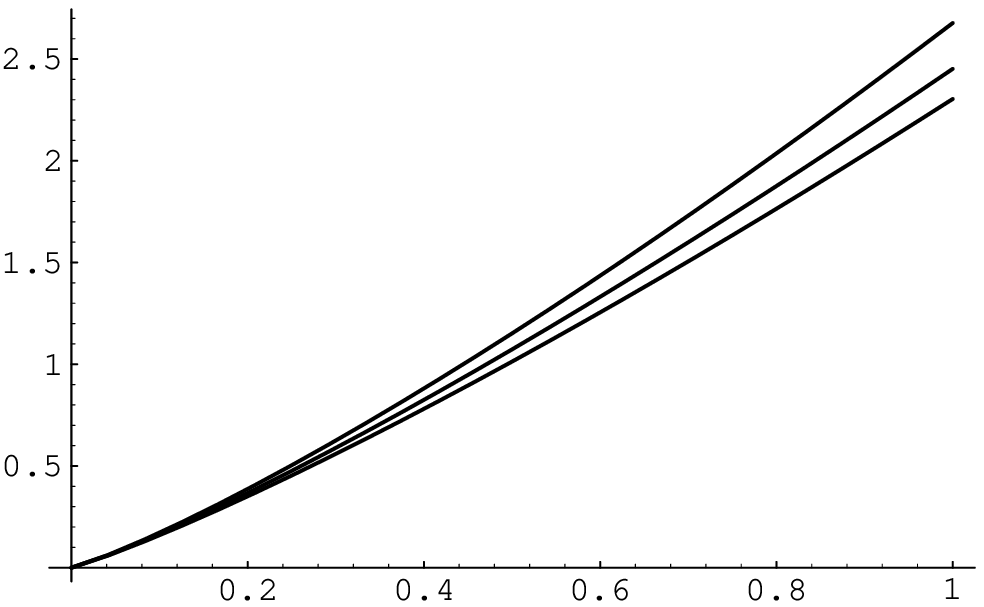}{\fer}{Fermion masses (\ref{fivexvi})
(top and bottom) 
and photon mass (\ref{fiveviii}) (middle) in softly broken 
$SU(2)$ Yang-Mills, 
as a function of $0 \le f_0 \le 1$.}{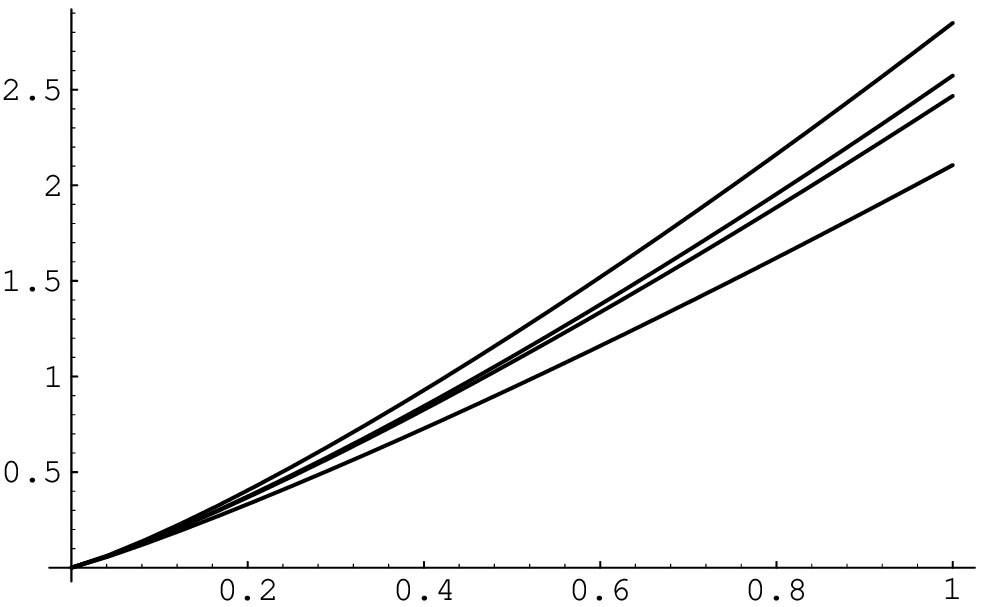}{\sca}{Masses of the
 scalars in softly broken 
$SU(2)$ Yang-Mills, as a function of $0 \le f_0\le 1$.}

These derivatives diverge at the monopole singularity 
$u=1$, and we may think that this can give some kind of singular behaviour 
there. In fact this is not so. The position of 
the minimum, $u_0$, behaves almost 
linearly with respect to $f_0$, $u_0-1 \sim f_0$, and this guarantees that 
the behaviour very near to $u=1$ (corresponding to a very small $f_0$) is 
perfectly smooth, as one can see in the figures. In \fig{\fer} we plot 
the fermion masses (\ref{fivexvi}) (top and bottom) and the photon mass 
given in (\ref{fiveviii}) (middle). In \fig{\sca} we plot the masses of 
the scalars, where the second one from the top corresponds to 
the doubly degenerated eigenvalue (\ref{fivexviii}).

\section*{Acknowledgments}

One of us (L. A.-G.) would like to thank J.M. Drouffe and 
J.B. Zuber for the opportunity to present this work at
the conference in honour of C. Itzykson ``The Mathematical
Beauty of Physics".  We would also like to thank 
J. Distler and C. Kounnas for an enjoyable collaboration.


\section*{References}
\begin{thebibliography}{99}


\bibitem{swone}
N. Seiberg and E. Witten, Nucl. Phys. {\bf B426} (1994) 19, hep-th/9407087.

\bibitem{swtwo}
N. Seiberg and E. Witten, Nucl. Phys. {\bf B431} (1994) 484,
hep-th/9408099.

\bibitem{kl}
A. Klemm, W. Lerche. S. Theisen and S. Yankielowicz, Phys. Lett. {\bf B344}
 (1995) 169, hep-th/9411048;\\
P.C. Argyres and A.E. Faraggi, Phys. Rev. Lett. {\bf 74} (1995) 3931,
hep-th/9411057.

\bibitem{klt}
A. Klemm, W. Lerche. and S. Theisen, Int. J. Mod. Phys. 
{\bf A11} (1996) 1929, hep-th/9505150.

\bibitem{ds}
M. Douglas and S.H. Shenker, Nucl. Phys. {\bf B447} (1995) 271,
hep-th/9503163.

\bibitem{groups}
A. Brandhuber and K. Landsteiner, Phys. Lett. {\bf B358} (1995) 73,
hep-th/9507008;\\
U.H. Danielsson and B. Sundborg, Phys. Lett. {\bf B358}
(1995) 273, hep-th/9504102;\\
A. Hanany and Y. Oz, Nucl. Phys. {\bf B452} (1995) 283, hep-th/9505075;\\
P.C. Argyres, M.R. Plesser and A.D. Shapere, Phys. Rev. Lett. {\bf 75}
(1995) 1699, hep-th/9505100.\quad

\bibitem{four}
B. de Wit and A. Van Pr\oe yen, Nucl. Phys. {\bf B245} (1984) 89;\\
see also P. Fr\'e and P. Soriani, ``The $N=2$ Wonderland",
 World Scientific, 1995, for a complete set of references.

\bibitem{five}
G. `t Hooft, 1976, in ``High Energy Physics", edited by A. Zichichi,
Palermo, 1976;\\
S. Mandelstam, Phys. Rep. {\bf C23} (1976) 245.


\bibitem{none}
I. Affleck, M. Dine and N. Seiberg, Nucl. Phys. {\bf B241} (1984) 493;
 {\bf B256} (1985) 557;\\
D. Amati, G.C. Rossi, G. Veneziano, Nucl. Phys. {\bf B249} (1985) 1;
D. Amati, K. Konishi, Y. Meurice, G.C. Rossi and G. Veneziano, Phys.
Rep. {\bf 162} (1988) 169;\\
T.R. Taylor, G. Veneziano and S. Yankielowicz,  Nucl. Phys. {\bf B218} (1982);
G. Veneziano and S. Yankielowicz, Phys. Lett. {\bf 113B} (1982) 231;\\  
N. Seiberg, Phys. Lett. {\bf B318} (1993) 469, hep-ph/9309335;
Phys. Rev. {\bf D49} (1994) 6857, hep-th/9402044;\\
K. Intriligator, R. Leigh and N. Seiberg, Phys. Rev. {\bf D50} (1994)
1052, hep-th/9403198;\\
K. Intriligator, Phys. Lett. {\bf B336} (1994) 409, hep-th/9407106;\\
K. Intriligator and N. Seiberg, Nucl. Phys. {\bf B431} (1994) 551,
hep-th/9408155.

\bibitem{nadual}
N. Seiberg, Nucl. Phys. {\bf B435} (1995) 129,  hep-th/9411149;\\
P.C. Argyres, M.R. Plesser, N. Seiberg and E. Witten,  Nucl. Phys. {\bf B461}
(1996) 71, hep-th/9511154.

\bibitem{girar}
L. Girardello and M.T. Grisaru, Nucl. Phys. {\bf B194} (1982) 65.

\bibitem{softone}
O. Aharony, J. Sonnenschein, M.E. Peskin and S. Yankielowicz, Phys.
Rev. {\bf D52} (1995) 6157, hep-th/9507013.

\bibitem{softwo}
N. Evans, S.D.H. Hsu and M. Schwetz, Phys. Lett. {\bf B355} (1995)
475, hep-th/9503186;\\
N. Evans, S.D.H. Hsu, M. Schwetz, S.B. Selipsky, Nucl. Phys. {\bf B456}
(1995) 205, hep-th/9508002.

\bibitem{XI}
S. Kachru and C. Vafa, Nucl. Phys. {\bf B450} (1995) 69, hep-th/9505105.

\bibitem{XII}
S. Kachru, A. Klemm, W. Lerche, P. Mayr, and C. Vafa, Nucl. Phys.
{\bf B459} (1996) 537, hep-th/9508155.


\bibitem{XIII}
B. de Wit, hep-th/9602060.

\bibitem{ad}
P.C. Argyres and M. Douglas, Nucl. Phys. {\bf B448} (1995) 166,
hep-th/9505062.


\bibitem{matone}
M. Matone, Phys. Lett. {\bf B357} (1995) 342, hep-th/9506102.

\bibitem{sonn}
J. Sonnenschein, S. Theisen and S. Yankielowicz, Phys. Lett. {\bf B367}
(1996) 145, hep-th/9510129.

\bibitem{ey}
T. Eguchi and S.-K. Yang, Mod. Phys. Lett. {\bf A11} (1996) 131,
 hep-th/9510183.

\bibitem{soft}
L. \'Alvarez-Gaum\'e, J. Distler, C. Kounnas and M. Mari\~no, hep-th/9604004.

\bibitem{XIV}
M.K. Prasad and C.M. Sommerfield, Phys. Rev. Lett. {\bf 35} (1975)
760;\\
E.B. Bogomolny, Sov. J. Nucl. Phys. {\bf 24} (1976) 449.

\bibitem{XV}
E. Witten and D. Olive, Phys. Lett. {\bf B78} (1978) 97.

\bibitem{XVI}
I.S. Gradshteyn and I.M. Ryzhik, ``Tables of series, products and integrals",
Academic Press.

\bibitem{dyon}
E. Witten, Phys. Lett. {\bf B86} (1979) 283.

\bibitem{agmm}
L. \'Alvarez-Gaum\'e and M. Mari\~no, to appear.

\bibitem{mass}
S. Ferrara, L. Girardello and F. Palumbo, Phys. Rev. {\bf D20} (1979) 403.



\end{thebibliography}
\end{document}